\documentclass[prc,aps,eqsecnum,floatfix,nofootinbib]{revtex4-1}
\usepackage{ulem}  
\usepackage{dcolumn}
\usepackage{bm}
\usepackage{color}
\usepackage{amssymb}
\usepackage{amsmath}
\usepackage{graphicx}
\usepackage{amsfonts}
\usepackage{slashed}
\usepackage{pstricks}
\usepackage{float}
\usepackage{hyperref}
\usepackage{array}
\usepackage{epsf}
\usepackage{soul}
\usepackage{siunitx}
\allowdisplaybreaks{
\begin{document}
\title{Local chiral interactions and magnetic structure of few-nucleon systems}
\author{R.\ Schiavilla$^{\rm a,b}$, A.\ Baroni$^{\rm c}$, S.\ Pastore$^{\rm d}$, M.\ Piarulli$^{\rm d}$,
L.\ Girlanda$^{\rm e,f}$, A.\ Kievsky$^{\rm g}$, A.\ Lovato$^{\rm h,i}$, L.E.\ Marcucci$^{\rm j,g}$, 
Steven C.\ Pieper$^{\rm h}$, M.\ Viviani$^{\rm g}$, and R.B.\ Wiringa$^{\rm h}$}
\affiliation{
$^{\rm a}$\mbox{Department of Physics, Old Dominion University, Norfolk, Virginia 23529, USA}\\
$^{\rm b}$\mbox{Theory Center, Jefferson Lab, Newport News, Virginia 23606, USA}\\
$^{\rm c}$\mbox{Department of Physics, University of South Carolina, Columbia, South Carolina 29208, USA}\\
$^{\rm d}$\mbox{Department of Physics, Washington University in St.~Louis, St.~Louis, Missouri 63130, USA}\\
$^{\rm e}$\mbox{Department of Mathematics and Physics, University of Salento, 73100 Lecce, Italy} \\
$^{\rm f}$\mbox{INFN-Lecce, 73100 Lecce, Italy} \\
$^{\rm g}$\mbox{INFN-Pisa, 56127 Pisa, Italy}\\
$^{\rm h}$\mbox{Physics Division, Argonne National Laboratory, Argonne, Illinois 60439, USA}\\
$^{\rm i}$\mbox{INFN-TIFPA, Trento Institute for Fundamental Physics and Applications, 38123 Trento, Italy}\\
$^{\rm j}$\mbox{Department of Physics, University of Pisa, 56127 Pisa, Italy}\\
}
\date{\today}
\begin{abstract}
The magnetic form factors of $^2$H, $^3$H, and $^3$He, deuteron photodisintegration cross sections
at low energies, and deuteron threshold electrodisintegration cross sections at backward angles in a wide
range of momentum transfers, are calculated with the chiral two-nucleon (and three-nucleon) interactions
including $\Delta$ intermediate states that have recently been constructed in configuration space.  The
$A\,$=$\,$3 wave functions are obtained from hyperspherical-harmonics solutions of the Schr\"odinger
equation.  The electromagnetic current includes one- and two-body terms, the latter induced by one- and
two-pion exchange (OPE and TPE, respectively) mechanisms and contact interactions.  The contributions
associated with $\Delta$ intermediate states are only retained at the OPE level, and are neglected in TPE
loop (tree-level) corrections to two-body (three-body) current operators.  Expressions for these currents
are derived and regularized in configuration space for consistency with the interactions.  The low-energy
constants that enter the contact 
few-nucleon systems.  The predicted form factors and deuteron electrodisintegration
cross section are in excellent agreement with experiment for momentum transfers up to 2--3 fm$^{-1}$.
However, the experimental values for the deuteron photodisintegration cross section are consistently
underestimated by theory, unless use is made of the Siegert form of the electric dipole transition operator.
A complete analysis of the results is provided, including the clarification of the origin of the aforementioned
discrepancy.
\end{abstract}

\index{}\maketitle
\section{Introduction}
\label{sec:intro}
The last few years have seen the development of chiral two-nucleon ($2N$) interactions
that are local in configuration space~\cite{Gezerlis:2013,Piarulli:2015,Piarulli:2016} and
therefore well suited for use in quantum Monte Carlo (QMC) calculations of light-nuclei
spectra and neutron-matter properties~\cite{Gezerlis:2014,Lynn:2014,Lynn:2016,Tews:2016,
Gandolfi:2017,Lynn:2017,Piarulli:2018}.  In conjunction with these, chiral (and local)
three-nucleon ($3N$) interactions have also been constructed~\cite{Epelbaum:2002,Piarulli:2018},
and the low-energy constants (LECs)
that characterize their contact terms---the LECs $c_D$ and $c_E$---have been constrained
either by fitting exclusively strong-interaction observables~\cite{Lynn:2016,Tews:2016,Lynn:2017,
Piarulli:2018} or by relying on a combination
of strong- and weak-interaction ones~\cite{Gazit:2009,Marcucci:2012,Baroni:2018}.  This last approach is made possible
by the relation, established in $\chi$EFT~\cite{Gardestig:2006}, between $c_D$ in the three-nucleon
contact interaction and the LEC in the $2N$ contact axial current~\cite{Gazit:2009,Marcucci:2012,Schiavilla:2017},
which allows one to use nuclear properties governed by either the strong  or weak interactions to
constrain simultaneously the $3N$ interaction and $2N$ axial current.

In the present study we adopt the $2N$ and $3N$ interactions constructed by
our group~\cite{Piarulli:2015,Piarulli:2016,Piarulli:2018,Baroni:2018}. The $2N$
interactions consist of an electromagnetic-interaction component, including up to
quadratic terms in the fine-structure constant, and a strong-interaction component
characterized by long- and short-range parts~\cite{Piarulli:2015,Piarulli:2016}.  The
long-range part retains one- and two-pion exchange (respectively, OPE and TPE) terms
from leading and sub-leading $\pi N$~\cite{Fettes:2000} and $\pi N\Delta$~\cite{Krebs:2007}
chiral Lagrangians up to next-to-next-leading order (N2LO) in the low-energy expansion.
In coordinate space, this long-range part is represented by charge-independent central, spin,
and tensor components with and without isospin-dependence ${\bm \tau}_i\cdot{\bm \tau}_j$
(the so-called $v_6$ operator structure), and by central and tensor components induced by
OPE and proportional to the isotensor operator $T_{ij}\,$=$\,3\,\tau_{i,z}\, \tau_{j,z}-{\bm \tau}_i\cdot{\bm \tau}_j$.
The radial functions multiplying these operators are singular at the origin, and are regularized
by a coordinate space cutoff of the form given in Eq.~(\ref{eq:cfl}) below.
The short-range part is described by charge-independent contact interactions up to N3LO,
specified by a total of 20 LECs, and charge-dependent ones up to NLO, characterized
by 6 LECs~\cite{Piarulli:2016}. By utilizing Fierz identities, the resulting charge-independent
interaction can be made to contain, in addition to the $v_6$ operator structure, spin-orbit, ${\bf L}^2$ (${\bf L}$
is the relative orbital angular momentum), and quadratic spin-orbit components, while the
charge-dependent one retains central, tensor, and spin-orbit components.  Both are regularized
by multiplication of a Gaussian cutoff ~\cite{Piarulli:2016}.

Two classes of interactions were constructed, which only differ in the range of laboratory energy
over which the fits to the {$2N$} database~\cite{Perez:2013} were carried out, either 0--125 MeV
in class I or 0--200 MeV in class II.  For each class, three different sets of cutoff radii
$(R_{\rm S},R_{\rm L})$ were considered $(R_{\rm S},R_{\rm L})\,$=$\,(0.8,1.2)$ fm in set a, (0.7,1.0)
fm in set b, and (0.6,0.8) fm in set c.  The $\chi^2$/datum achieved by the fits in class I (II) was
$\lesssim 1.1$ $(\lesssim1.4)$ for a total of about 2700 (3700) data points.  We have been referring to these
high-quality {$2N$} interactions generically as the Norfolk $v_{ij}$'s (NV2s), and have been designating those
in class I as NV2-Ia, NV2-Ib, and NV2-Ic, and those in class II as NV2-IIa, NV2-IIb, and NV2-IIc.
Owing to the poor convergence of the hyperspherical-harmonics (HH) expansion and the severe
fermion-sign problem of the Green's function Monte Carlo (GFMC) method, however,
models Ic and IIc have not been used so far in actual calculations of light nuclei.

The $3N$ interactions consist~\cite{Epelbaum:2002} of a long-range piece mediated by TPE, including
$\Delta$ intermediate states~\cite{Piarulli:2018}, at LO and NLO, and a short-range piece parametrized
in terms of two contact interactions, which enter formally at NLO, proportional to the LECs $c_D$ and $c_E$.
Two distinct sets were constructed.  In the first, $c_D$ and $c_E$ were determined
by simultaneously reproducing the experimental trinucleon ground-state energies and $nd$ doublet
scattering length for each of the $2N$ models considered, namely NV2-Ia/b and NV2-IIa/b~\cite{Piarulli:2018}.
In the second set, these LECs were constrained by fitting, in addition to the trinucleon energies, the
empirical value of the Gamow-Teller matrix element in tritium $\beta$ decay~\cite{Baroni:2018}.
The resulting Hamiltonian models were designated as NV2+3-Ia/b and NV2+3-IIa/b (or Ia/b and IIa/b
for short) in the first case, and as NV2+3-Ia$^*$/b$^*$ and NV2+3-IIa$^*$/b$^*$ (or Ia$^*$/b$^*$
and IIa$^*$/b$^*$) in the second.  These two different procedures for fixing $c_D$ and $c_E$ produced
rather different values for these LECs\footnote{It is worthwhile observing here that the strong
correlation between $^3$H/$^3$He binding energies and $nd$ doublet scattering length makes the
determination of $c_D$ and $c_E$ somewhat problematic in Ref.~\cite{Piarulli:2018}.  This difficulty is
removed in Ref.~\cite{Baroni:2018}.}, particularly for $c_E$ which was found to be relatively large
and negative in models Ia/b and IIa/b, but quite small, and not consistently negative, in models Ia$^*$/b$^*$
and IIa$^*$/b$^*$.  This in turn impacts predictions for the spectra of light nuclei and the equation of
state of neutron matter, since a negative $c_E$ leads to repulsion in light nuclei, but attraction in
neutron matter. In particular, while model Ia provides an excellent description of the energy
levels and level ordering of nuclei in the mass range $A\,$=$\,$4--12~\cite{Piarulli:2018}, it collapses
neutron matter already at relatively low densities~\cite{Piarulli:2018a}, and cannot sustain the existence
of neutron stars of twice solar masses, in conflict with recent observations~\cite{Demorest:2010,Antoniadis:2013}.
By contrast, there are indications that the smaller values of $c_E$ characteristic of models
Ia$^*$/b$^*$ and IIa$^*$/b$^*$, mitigate, if not resolve, the collapse problem, while still predicting
light-nuclei spectra in reasonable agreement with experimental data~\cite{Piarulli:2018a}.

Electromagnetic properties of few-nucleon systems are among the observables of choice for testing
models of nuclear interactions and associated electromagnetic charge and current operators.\footnote{In
this connection, the first-principles calculation of magnetic moments of few-nucleon systems in lattice
quantum chromodynamics reported recently by the NPLQCD Collaboration~\cite{Beane:2014} should be noted.}
Nuclear electromagnetic charge and current operators in a $\chi$EFT formulation with nucleon
and pion degrees of freedom were derived up to one loop originally by Park {\it et al.}~\cite{Park:1993,Park:1996}
in covariant perturbation theory.  Subsequently, two independent derivations, based on time-ordered
perturbation theory (TOPT), appeared in the literature, one by some of the present
authors~\cite{Pastore:2009,Pastore:2011, Piarulli:2013} and the other by
K\"olling {\it et al.}~\cite{Koelling:2009,Koelling:2011}.  These two derivations differ in the way in which
non-iterative terms are isolated in reducible contributions\footnote{In the pioneering work of Park
{\it et al.}~\cite{Park:1996} only irreducible contributions were retained.}. The authors of
Refs.~\cite{Koelling:2009,Koelling:2011} use TOPT in combination with the unitary
transformation method~\cite{Okubo:1954} to decouple, in the Hilbert space of pions and nucleons,
the states consisting of nucleons only from those including, in addition, pions.  In contrast, we
construct an interaction such that, when iterated in the Lippmann-Schwinger equation, generates a
$T$-matrix matching, order by order in the power counting, the $\chi$EFT amplitude calculated in
TOPT~\cite{Pastore:2011,Pastore:2008}.  These two different formulations lead to the same
formal expressions for the electromagnetic current operator up to one loop (or N3LO).  However,
some differences remain in the electromagnetic charge operator, specifically in some of the
pion-loop corrections to its short-range part~\cite{Piarulli:2013}.  They are not relevant here (and
will not be discussed any further), since we are primarily interested in magnetic structure and
response.

A partial derivation of the electromagnetic current in a $\chi$EFT formulation
which explicitly accounts for $\Delta$ intermediate states in TPE contributions
was carried out in Ref.~\cite{Pastore:2008}.  However, a systematic study of
these contributions in two-body (as well as three-body) currents is not yet available
(within $\chi$EFT).  In the present work, we only retain $\Delta$ contributions at the OPE
level, which formally enter at N2LO in the chiral expansion, and ignore altogether their
contributions to TPE mechanisms at N3LO.  There are indications from an earlier
study~\cite{Marcucci:1998} which approximately accounted for explicit $\Delta$ components
in nuclear ground states with the transition-correlation-operator method~\cite{Schiavilla:1992},
that the latter are much smaller than the former in the low-momentum transfer region of the
trinucleon magnetic form factors of interest here (see Figs.~13--14 and 18--19 of
Ref.~\cite{Marcucci:1998}).  Thus, we do not expect the present incomplete
treatment of $\Delta$ effects to significantly
affect our predictions for the two- and three-body observables we consider here.

This paper is organized as follows.  In Sec.~\ref{sec:em3} we list explicitly
the configuration-space expressions for the electromagnetic current up to N3LO.
Those up to N2LO are well known, and are reported here for completeness and
clarity of presentation.  However, the configuration-space
expressions for the loop corrections at N3LO were, to the best of our knowledge,
not previously known; they are derived in Appendix~\ref{app:a1} of the present paper.  In Sec.~\ref{sec:fit}
we determine the unknown LECs that enter the current at N3LO by fitting the magnetic
moments of $^2$H, $^3$H, and $^3$He for each of the Hamiltonian models considered
(Ia$^*$/b$^*$ and IIa$^*$/b$^*$), and in Sec.~\ref{sec:res} present predictions for the
magnetic form factors of these nuclei, the deuteron photodisintegration cross section
for photon energies ranging from threshold up to 30 MeV, and the deuteron threshold
electrodisintegration cross section at backward angles for momentum transfers up to
about 5.5 fm$^{-1}$ corresponding to these models (as well as models Ia/b and IIa/b for the
trinucleon form factors), along with a fairly detailed analysis of these results.  Finally,
we offer some concluding remarks in Sec.~\ref{sec:concl}.

\section{Electromagnetic current up to N3LO}
\label{sec:em3}
\begin{figure}[bth]
\includegraphics[width=1.5in]{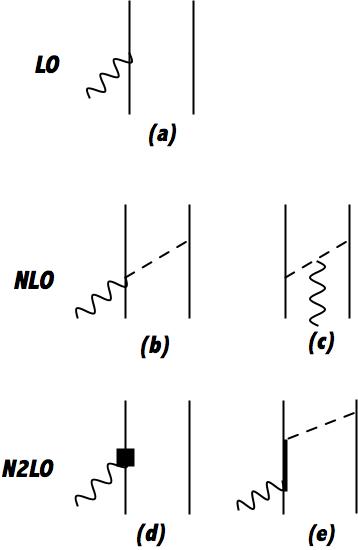}
\caption{Diagrams illustrating the contributions to the electromagnetic current up
to N2LO (with power scaling up to $Q^{0}$).  Nucleons, $\Delta$-isobars, pions, and external
fields are denoted by solid, thick-solid, dashed, and wavy lines,
respectively.  The square in panel (d) represents relativistic
corrections to the LO current.  Only a single time ordering is shown in panels (b), (c), and (e).}
\label{fig:f1a}
\end{figure}

\begin{figure}[bth]
\includegraphics[width=2.75in]{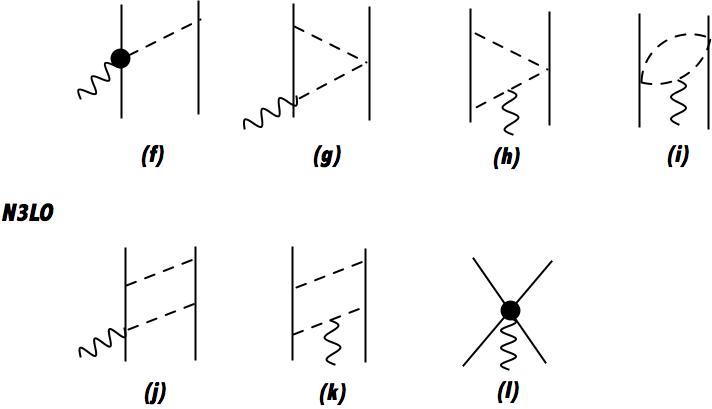}
\caption{Diagrams illustrating the contributions to the electromagnetic current at
N3LO (with power scaling $Q$) without the inclusion of $\Delta$ intermediate states.
Nucleons, pions, and external fields are denoted by solid, dashed, and wavy lines,
respectively.  The solid circle in panel (f) is associated with $\gamma\pi N$ interaction
vertices generated by ${\cal L}^{(3)}_{\pi N}$~\cite{Fettes:2000}.}
\label{fig:f1b}
\end{figure}

We illustrate the contributions to the two-body electromagnetic current 
in a $\chi$EFT with nucleon, $\Delta$-isobar, and pion degrees of
freedom up to N2LO in Fig.~\ref{fig:f1a}, and the contributions at N3LO
excluding $\Delta$ intermediate states in Fig.~\ref{fig:f1b}.  They
have been derived in a number of papers in approaches based on
either covariant perturbation theory~\cite{Park:1993,Park:1996} or, more
recently, time-ordered perturbation
theory~\cite{Pastore:2009,Koelling:2009,Pastore:2011,Koelling:2011,Piarulli:2013}.
For completeness and ease of presentation, we report below the configuration-space
expressions of these various terms. 
The LO one in panel (a), which scales as $Q^{-2}$ in the power counting ($Q$ denotes generically
a low-momentum scale), reads
\begin{equation}
\label{eq:jlo}
{\bf j}^{\rm LO}({\bf q})=\frac{\epsilon_i }{2\, m}\,
 \left\{ {\bf p}_i\,\, ,\, \, {\rm e}^{i{\bf q}\cdot{\bf r}_i}  \right\} 
+i\,\frac{\mu_{i}}{2\, m} \,\, {\rm e}^{i{\bf q}\cdot{\bf r}_i}\,\, {\bm \sigma}_i\times {\bf q }\,
+ ( i \rightleftharpoons j)\ ,
 \end{equation}
where ${\bf p}_k\,$=$\, -i\, {\bm \nabla}_k$ and ${\bm \sigma}_k$
are the momentum and Pauli spin operators of
nucleon $k$, $m$ is its mass ($m\,$=$\,$938.9 MeV),
${\bf q}$ is the external field momentum, and we have defined
the isospin operators
\begin{eqnarray}
\epsilon_k&=& (1+\tau_{k,z})/2 \ , \qquad \mu_k=(\mu^S+\mu^V \tau_{k,z})/2 \ .
\end{eqnarray}
Here $\mu^S$ and $\mu^V$ are the isoscalar and isovector combinations
of the nucleon magnetic moments ($\mu^S\,$=$\,0.8798$ n.m.~and $\mu^V\,$=$\,4.7059$ n.m.).
The NLO terms in panels (b) and (c) with scaling $Q^{-1}$ are written as
\begin{widetext}
\begin{eqnarray}
{\bf j}^{\rm NLO}({\bf q})&=&  \left[ {\rm e}^{i\,{\bf q}\cdot {\bf r}_i}\,
 \left({\bm \tau}_i \times {\bm \tau}_j\right)_z
\, I_0^\pi(\mu_{ij}) \, {\bm \sigma}_i \,\, {\bm \sigma}_j \cdot \hat{\bf r}_{ij}+\! (i \rightleftharpoons j) \right] 
+\, {\rm e}^{i\,{\bf q}\cdot {\bf R}_{ij}}\, \left({\bm \tau}_i \times {\bm \tau}_j\right)_z\, 
 \,{\bm \sigma}_i \cdot \left({\bm \nabla}^\mu_{ij} + i\, \frac{\bf q}{2\, m_\pi}\right) \nonumber\\
&&\times {\bm \sigma}_j \cdot \left({\bm \nabla}_{ij}^\mu -i\, \frac{\bf q}{2\, m_\pi}\right)
 {\bm \nabla}^\mu_{ij}\, L_0^\pi({\bm\mu}_{ij},{\bf q}) \ ,
\label{eq:nlor2}
\end{eqnarray}
\end{widetext}
where
\begin{equation}
{\bf r}_{ij}={\bf r}_i-{\bf r}_j\ , \qquad {\bf R}_{ij}=\left({\bf r}_i+{\bf r}_j\right)/2 \ ,
\qquad {\bm \mu}_{ij}=m_\pi\, {\bf r}_{ij} \ ,
\end{equation}
the gradients ${\bm \nabla}^\mu_{ij}$ are
relative to the adimensional variables ${\bm \mu}_{ij}$, and the correlation functions
$I_0^\pi (\mu)$  and $L_0^\pi({\bm\mu},{\bf q})$ are defined as
\begin{eqnarray}
\label{eq:e12}
I_0^\pi (\mu )&=& -\frac{g^2_A}{16\, \pi}\, \frac{m_\pi^2}{f_\pi^2}\, (1+\mu )\, \frac{e^{-\mu}}{\mu^2} \ ,\\
\label{eq:e11}
L_0^{\pi}({\bm\mu},{\bf q}) &=&\frac{g^2_A}{16\, \pi}\, \frac{m_\pi^2}{f_\pi^2}
\int_{-1/2}^{1/2} dz\,\,
 {\rm e}^{-i\,z\left({\bf q}/m_\pi\right) \cdot {\bm \mu}} \,\, \frac{e^{-\mu \, L(z,q)}}{L(z,q)} \ , 
\end{eqnarray}
with
\begin{equation}
 L(z,q)=\sqrt{1+\frac{q^2}{4\, m_\pi^2}\left(1-4\, z^2\right)} \ ,
 \end{equation}
and $g_A$ and $f_\pi$ are, respectively, the nucleon axial coupling constant
($g_A\,$=$\, 1.29$) and pion-decay amplitude ($f_\pi\,$=$\, 92.4$ MeV),
and $m_\pi$ is average pion mass ($m_\pi\,$=$\,$138.039 MeV) (these values
are taken from Ref.~\cite{Piarulli:2015}).  The N2LO
terms (with scaling $Q^0$) consist of relativistic corrections (RC) to the LO
current, panel (d), and contributions involving $\Delta$ intermediate states
($\Delta$), panel (e),
\begin{widetext}
\begin{eqnarray}
{\bf j}^{\rm N2LO}_{\rm RC}({\bf q})&=&-\frac{\epsilon_i}{16 \, m^3}
 \left\{2 \left( p_i^2 +\frac{q^2}{4} \right)  \big( 2\, {\bf p}_i
+i\, {\bm \sigma}_i\times {\bf q } \big) 
+ {\bf p}_i\cdot {\bf q}\, \left({\bf q} +2\, i\, {\bm \sigma}_i\times {\bf p }_i \right)
 \,\, ,\,\, 
 {\rm e}^{i {\bf q}\cdot {\bf r}_i}\right\}\nonumber \\
 && \hspace{-0.2cm} - i\, \frac{\mu_i-\epsilon_i}{16 \, m^3}
  \left\{ {\bf p}_i\cdot {\bf q} \big( 4\, {\bm \sigma}_i\times {\bf p}_i-i\, {\bf q}\big) 
 - \left(  2\, i\, {\bf p}_i -{\bm \sigma}_i\times {\bf q} \right)\frac{q^2}{2}  +2\, \left({\bf p}_i\times {\bf q}\right)
 \, {\bm \sigma}_i\cdot {\bf p}_i
 \,\, , \,\, {\rm e}^{i {\bf q}\cdot {\bf r}_i}\right\} + (i \rightleftharpoons j)  \ ,
 \label{eq:j1rcc}\\
 \label{eq:jdn2lo}
{\bf j}^{\rm N2LO}_{\Delta}({\bf q})&=&-i\,{\rm e}^{i {\bf q}\cdot {\bf r}_i} \, \tau_{j,z}
\left[ I^{\Delta}_1(\mu_{ij}) \, {\bm \sigma}_j
+I_{2}^{\Delta}(\mu_{ij}) \,\,
 {\bm \sigma}_j\cdot \hat{\bf r}_{ij} \,\, \hat{\bf r}_{ij}\right]\times 
 \frac{{\bf q}}{m_\pi} \nonumber\\
&&+ \frac{i}{4}\, {\rm e}^{i\, {\bf q}\cdot {\bf r}_i}\, \left({\bm \tau}_i\times{\bm \tau}_j\right)_z \left[
I^{\Delta}_1 (\mu_{ij})\, {\bm \sigma}_i \times {\bm \sigma}_j
+I^\Delta_2(\mu_{ij})\, {\bm \sigma}_j\cdot \hat{\bf r}_{ij}\,\, {\bm \sigma}_i\times \hat{\bf r}_{ij}\right] \times\frac{{\bf q}}{m_\pi} +(i\rightleftharpoons j )\ ,
\end{eqnarray}
\end{widetext}
where the correlation functions $I^\Delta_k(\mu)$ are
 \begin{eqnarray}
\label{eq:e13}
I^\Delta_1(\mu)&=&- \left(\frac{g_A\,h_A}{18\, \pi}\,
\frac{\mu_{\Delta N}}{2\, m}\, \frac{m_\pi^2}{m_{\Delta N}}
 \frac{m_\pi^2}{f_\pi^2}\right) (1+\mu )\, \frac{e^{-\mu}}{\mu^3} \ , \\
\label{eq:e14}
I^\Delta_2(\mu)&=&\left(\frac{g_A\,h_A}{18\, \pi}\,
\frac{\mu_{\Delta N}}{2\, m}\, \frac{m_\pi^2}{m_{\Delta N}}
 \frac{m_\pi^2}{f_\pi^2}\right)
 (3+3\,\mu+\mu^2)\, \frac{e^{-\mu}}{\mu^3} \ , 
\end{eqnarray}
and $h_A$ and $\mu_{\Delta N}$ are, respectively, the nucleon-to-$\Delta$
transition axial coupling constant  ($h_A\,$=$\,$2.74) and magnetic moment
($\mu_{\Delta N}\,$=$\,$3 n.m.~\cite{Carlson:1986}), and $m_{\Delta N}$ is
$\Delta$-nucleon mass difference ($m_{\Delta N}\,$=$\,$293.1 MeV).
The N3LO terms are written as the sum of an isoscalar OPE contribution, panel (f),
\begin{equation}
{\bf j}^{\rm N3LO}_{\rm OPE}({\bf q})=-i\,{\rm e}^{i {\bf q}\cdot {\bf r}_i} \,  {\bm \tau}_i\cdot{\bm \tau}_j
\left[ I^\pi_1(\mu_{ij}) \, {\bm \sigma}_j
+I^\pi_2(\mu_{ij}) \,\,
 {\bm \sigma}_j\cdot \hat{\bf r}_{ij} \,\, \hat{\bf r}_{ij}\right]\times 
 \frac{{\bf q}}{m_\pi}+(i\rightleftharpoons j )\ ,
 \label{eq:e212}
\end{equation}
isovector TPE contributions, panels (g)-(k),
\begin{eqnarray}
\label{eq:jloop}
{\bf j}^{{\rm N3LO}}_{\rm TPE}({\bf q})&=&
i\,\tau_{j,z}\,\, {\rm e}^{i{\bf q}\cdot {\bf R}_{ij}}
\left\{ 
\left[ F^{(0)}_0(\lambda_{ij})  + F^{(1)}_2(\lambda_{ij})\right]{\bm \sigma}_i 
+ F^{(2)}_2(\lambda_{ij})\,   {\bm \sigma}_i\cdot\hat{\bf r}_{ij} \,\, \hat{\bf r}_{ij}
 \right\} \times \frac{{\bf q}}{2\, m_\pi}  \nonumber\\
&&\hspace{-0.2cm}-\frac{1}{2} ({\bm \tau}_i\times {\bm \tau}_j)_z\,\,  {\rm e}^{i{\bf q}\cdot {\bf R}_{ij}} \,\hat{\bf r}_{ij}\,
\left[ \lambda_{ij}\, \frac{v^{\rm NLO}_{\,2\pi}(\lambda_{ij})}{2\,m_\pi} \right]
+ (i \rightleftharpoons j) \ , 
\end{eqnarray}
and both isoscalar and isovector contact contributions, panel (l),
from minimal (MIN) and non-minimal (NM) couplings and from the
regularization scheme in configuration space 
we have adopted for the TPE current  (and labeled CT) in Appendix~\ref{app:a1}, respectively,
\begin{widetext}
\begin{eqnarray}
\label{eq:jmin}
{\bf j}^{\rm N3LO}_{\rm MIN}({\bf q})&=&\frac{1 }{8}\,  \,
\left({\bm \tau}_i\times{\bm \tau}_j\right)_z\,\,
{\rm e}^{i{\bf q}\cdot{\bf R}_{ij}}\
C^{(1)}_{R_{\rm S}}(z_{ij}) 
\Big[ \, m_\pi^4\left( C_2+3\, C_4+C_7\right) \hat {\bf r}_{ij} 
+ m_\pi^4\left( C_2-C_4-C_7\right)  \,\hat {\bf r}_{ij} \, \,{\bm \sigma}_i\cdot{\bm \sigma}_j\nonumber\\
&&\hspace{-0.2cm}+ \, m_\pi^4\, C_7 \left( {\bm \sigma}_i\cdot \hat{\bf r}_{ij}\,\, {\bm \sigma}_j
+ {\bm \sigma}_j\cdot \hat{\bf r}_{ij}\,\, {\bm \sigma}_i\right) \Big]
-\frac{1}{8} m_\pi^4\, C_5 \,{\rm e}^{i{\bf q}\cdot{\bf R}_{ij}} \Big[ \left(\tau_{i,z}-\tau_{j,z}\right)\, 
C^{(1)}_{R_{\rm S}}(z_{ij}) 
 \left({\bm \sigma}_i+{\bm \sigma}_j\right)
\times \hat{\bf r}_{ij}  \nonumber\\
&&+\, i\, C^{(0)}_{R_{\rm S}}(z_{ij})\left({\bm \sigma}_i+{\bm \sigma}_j\right)
\times \frac{\bf q}{m_\pi}\Big] \ , \\
\label{eq:jnm}
{\bf j}^{\rm N3LO}_{\rm NM}({\bf q})&=& - i \, {\rm e}^{i{\bf q}\cdot{\bf R}_{ij}}\, 
C^{(0)}_{R_{\rm S}}(z_{ij})\Big[  m_\pi^4\, C_{15}^\prime 
\left({\bm \sigma}_i+{\bm \sigma}_j\right)
+  m_\pi^4\, C_{16}^\prime\, (\tau_{i,z} - \tau_{j,z})  \left({\bm \sigma}_i-{\bm \sigma}_j\right)
 \Big]\, \times \frac{{\bf q}}{m_\pi}  \ ,\\
 \label{eq:jloopct}
{\bf j}^{{\rm N3LO}}_{\rm CT}({\bf q})&=&
i\,\tau_{j,z}\,\, {\rm e}^{i{\bf q}\cdot {\bf R}_{ij}}
\,\, F^{(0)}_0(z_{ij};\infty)\,\, {\bm \sigma}_i  \times \frac{{\bf q}}{2\, m_\pi} 
+ (i \rightleftharpoons j) \ , 
\end{eqnarray}
\end{widetext}
where we have introduced the notation
\begin{equation}
\label{eq:e217}
\lambda_{ij}=2\, m_\pi\, r_{ij} \ , \qquad z_{ij}=r_{ij}/R_{\rm S} \ ,
\end{equation}
and in the contact terms the $\delta$-function has been smeared by
a Gaussian cutoff and hence
\begin{equation}
\label{eq:e221}
C^{(0)}_{R_{\rm S}}(z) = \frac{1}{\pi^{3/2}\, (m_\pi\, R_{\rm S})^3} \, {\rm e}^{-z^2} \ , \qquad
C^{(1)}_{R_{\rm S}}(z) = \frac{1}{m_\pi\, R_{\rm S}}\, \frac{d\,C^{(0)}_{R_{\rm S}}(z)}{ dz} \ .
\end{equation}
The correlation functions $I^\pi_1(\mu)$ and $I^\pi_2(\mu)$ in Eq.~(\ref{eq:e212}) are defined as
in Eqs.~(\ref{eq:e13}) and~(\ref{eq:e14}), but for the combination of
constants in front of those equations being replaced by
\begin{equation}
\left(\frac{g_A\,h_A}{18\, \pi}\,
\frac{\mu_{\Delta N}}{2\, m}\, \frac{m_\pi^2}{m_{\Delta N}}
 \frac{m_\pi^2}{f_\pi^2}\right) 
 \longrightarrow \left(\frac{g_A}{16\, \pi}\,
 \frac{m_\pi^2}{f_\pi^2}\, m_\pi^2\, d_9^\prime \right) \ .
\end{equation}
The derivation of the loop correlation functions $F_0^{(0)}(z_{ij};\infty)$, $F_0^{(0)}(\lambda)$,
$F_2^{(1)}(\lambda)$, and $F_2^{(2)}(\lambda)$ is somewhat involved and is relegated in
Appendix~\ref{app:a1}---the relevant equations, where these functions are defined, are
(\ref{eq:e28}), (\ref{eq:e55}), and (\ref{eq:e56})--(\ref{eq:e57}), respectively.

Several comments are now in order. First, we have not accounted for explicit $\Delta$
intermediate states in the corrections at N3LO.  These enter in loops in two-body
operators---indeed, a partial derivation of them was already given in
Ref.~\cite{Pastore:2008}---and at tree-level in three-body operators (note that there
are no such terms at N3LO in a $\chi$EFT with nucleon and pion degrees of freedom
only~\cite{Girlanda:2010}).  Indeed, contributions due to these higher-order $\Delta$ currents
have yet to be studied quantitatively in calculations of electro- and photo-nuclear observables.  In this
context, we also note that the isovector OPE current at N3LO from the Lagrangian
${\cal L}^{(3)}_{\pi N}$~\cite{Fettes:2000}, which depends on the LECs $d_{8}^\prime$
and $d_{21}^\prime$ in the notation of Ref.~\cite{Piarulli:2013}, has been
assumed here to be saturated by the three-level $\Delta$ current of panel (e).

Second, the longitudinal term in the loop corrections of Eq.~(\ref{eq:jloop}) involves
the TPE interaction $v_{2\pi,ij}^{\rm NLO}\!=\!v_{2\pi}^{\rm NLO}(r_{ij})\, {\bm \tau}_i\cdot{\bm \tau}_j$ which
one obtains at NLO~\cite{Pastore:2009,Piarulli:2013} (with nucleons and pions only).
However, the chiral interactions adopted in the present study
receive TPE contributions also from $\Delta$ intermediate states at both NLO and
N2LO~\cite{Piarulli:2016}; as a matter of fact, up to N2LO included, they have the following
operator structure~\cite{Piarulli:2015,Piarulli:2016}
\begin{equation}
\label{eq:vlr}
\widetilde{v}_{2\pi,ij}^{\,\rm N2LO}= \sum_{p=1}^{6} \widetilde{v}_{\,2\pi}^{\,(p)} (r_{ij}) \, O^{(p)}_{ij} \ , \qquad
O^{(p=1,\dots,6)}_{ij}=\left[{\bf 1}\, ,\, {\bm \sigma}_i\cdot {\bm \sigma}_j\, , \,S_{ij}\right]
\otimes\left[{\bf 1}\, ,\, {\bm \tau}_i\cdot {\bm \tau}_j\right] \ ,
\end{equation}
where $S_{ij}$ is the standard tensor operator.  Hereafter, we will make the replacement
\begin{equation}
v_{2\pi}^{\rm NLO}(r_{ij})\longrightarrow \widetilde{v}_{\,2\pi}^{\,(2)} (r_{ij}) +
\widetilde{v}_{\,2\pi}^{\,(4)} (r_{ij})\,\, {\bm \sigma}_i\cdot {\bm \sigma}_j
+\widetilde{v}_{\,2\pi}^{\,(6)} (r_{ij})\,\,S_{ij} 
\end{equation}
in the longitudinal term of Eq.~(\ref{eq:jloop}), where $\widetilde{v}_{\,2\pi}^{\,(p)} (r_{ij}) \, O^{(p)}_{ij}$
with $p$ even denote the isospin-dependent central,
spin-spin, and tensor components.  While such a replacement is not consistent from a power-counting
perspective, it ensures, nevertheless, that the resulting TPE current satisfies the
continuity equation (with the corresponding
interaction components) in the limit of small momentum transfers, see Appendix~\ref{app:a1}.

Third, in the contact terms the LECs $C_i$ are taken from Ref.~\cite{Piarulli:2016}, where they have
been determined by fits to $pp$ and $np$ cross sections and polarization observables, including
$pp$, $np$, and $nn$ scattering lengths and effective ranges.  The determination of
the LECs $C_{15}^\prime$ and $C_{16}^\prime$ in the non-minimal current, and $d_9^\prime$
in the isoscalar OPE current at N3LO, is discussed in the following section.

Finally, the OPE and TPE correlation functions as well as the OPE ones resulting
from the application of the gradients to $L_0^\pi(\mu)$, denoted generically
with $X(m_\pi r)$ below, are each regularized by
multiplication of a configuration-space cutoff as in the case of the local chiral
interactions of Refs.~\cite{Piarulli:2015,Piarulli:2016,Baroni:2018}, namely
\begin{equation}
X(m_\pi r) \longrightarrow  C_{R_{\rm L}}(r) \, X( m_\pi r) \ ,
\end{equation}
with
\begin{equation}
\label{eq:cfl}
C_{R_{\rm L}}(r)  = 1 - \frac{1}{ \left(r/R_{\rm L}\right)^p\, 
{\rm e}^{(r-R_{\rm L})/a_{\rm L}}+1} \ ,
\end{equation}
where $a_{\rm L}\,$=$\,R_{\rm L}/2$, and the exponent $p$ is taken as $p\,$=$\, 6$ for consistency
with the interactions (note that the correlation functions in the TPE currents behave
as ${\rm ln}\,r/r^3$ in the limit $r\rightarrow 0$).  
\section{Determination of low-energy constants}
\label{sec:fit}
As already mentioned, the LECs $C_i$, $i=1,\dots,7$, in
the minimal contact current are taken from fits to
nucleon-nucleon scattering data~\cite{Piarulli:2016}.  In reference to
the LECs entering the OPE and non-minimal contact
currents at N3LO, it is convenient to introduce the adimensional
set $d_i^{S,V}$ (in units of $m_\pi$) as
\begin{equation}
C_{15}^\prime=d_1^S/m_\pi^4  \ , \qquad d_9^\prime=d_2^S/m_\pi^2\ , \qquad
C_{16}^\prime=d_1^V/m_\pi^4 \ ,
\end{equation}
where the superscript $S$ or $V$ on the $d^{S,V}_i$ characterizes the isospin of the
associated operator, i.e.,~whether it is isoscalar or isovector.  The values of these LECs are
listed in Table~\ref{tb:tb1}: $d_1^S$ and $d_2^S$ have been fixed by reproducing
the experimental deuteron magnetic moment $\mu_d$ and isoscalar
combination $\mu_S$ of the trinucleon magnetic moments, while $d_1^V$ has been
determined by their isovector combination.  Naive power counting would indicate
that the values for these LECs are natural.  Indeed, since the NLO and N3LO non-minimal (NM)
contact contributions scale as\footnote{This
scaling directly follows from the momentum-space expressions of the currents, listed explicitly
in Ref.~\cite{Piarulli:2013}.} (hereafter, the low momentum scale $Q$ is assumed to be of order $m_\pi$)
\begin{equation}
{\rm NLO}\sim \frac{g_A^2}{4\, f_\pi^2}\, \frac{1}{m_\pi} \ , \qquad {\rm N3LO(NM)}\sim \frac{1}{m^3_\pi}  \, d_1^{S,V} \ ,
\end{equation}
and since the ratio ${\rm N3LO(NM)}/{\rm NLO}$ is expected to be suppressed by $m^2_\pi/\Lambda_\chi^2$,
where $\Lambda_\chi$ is the hard scale which we take as $\Lambda_\chi\,$$\sim$$\,$1 GeV, it follows that
\begin{equation}
 d_1^{S,V} \sim \frac{g_A^2}{4}\, \frac{m_\pi^2}{f_\pi^2} \, \frac{m_\pi^2}{\Lambda_\chi^2} \sim 0.018 \ .
\end{equation}
A similar argument leads to the expectation that the LEC in the OPE (isoscalar) contribution at N3LO
has a magnitude of the order 
\begin{equation}
d_2^S \sim g_A\, \frac{m_\pi^2}{\Lambda_\chi^2} \sim 0.025 \ .
\end{equation}
Both these values are not out of line with those reported in Table~\ref{tb:tb1}.
\begin{table}[bth]
\begin{tabular}{l ||S|S||S|S}
                         & Ia$^*$          &Ib$^*$   & IIa$^*$ & IIb$^*$\\
\hline
$d_1^S$  &-0.00999  & -0.02511   & -0.01170 &-0.04955 \\
$d_2^S$  &-0.06571   & -0.02384   & -0.04714 & -0.07947 \\
\hline
$d_1^V$  & -0.05120  & -0.03509 & -0.05128& -0.03880 \\
\hline
\end{tabular}
\caption{Adimensional values of the isoscalar and isovector LECs corresponding
to the nuclear Hamiltonians NV2+3-Ia$^*$/b$^*$ and NV2+3-IIa$^*$/b$^*$~\cite{Baroni:2018},
designated as Ia$^*$/b$^*$ and IIa$^*$/b$^*$ for brevity.}
\label{tb:tb1}
\end{table}

The calculations of the observables are
based on the (chiral) two-nucleon interactions of Ref.~\cite{Piarulli:2016} for the
deuteron, augmented by the (chiral) three-nucleon interactions developed in
Refs.~\cite{Piarulli:2018,Baroni:2018} for $^3$He/$^3$H, and use, for the
trinucleon case, wave functions obtained from hyperspherical-harmonics (HH) solutions of 
the Schr\"odinger equation with these interactions (see Ref.~\cite{Kievsky:2008} for a
review of HH methods). The corresponding Hamiltonians are denoted as NV2+3-Ia$^*$/b$^*$
and NV2+3-IIa$^*$/b$^*$ (or Ia$^*$/b$^*$ and  IIa$^*$/b$^*$ for short), where I or II,
and a or b, specify, respectively, the energy range
over which the fits to the two-nucleon database were carried out (for the
two-nucleon interactions~\cite{Piarulli:2016})---either 0--125 MeV (I) or 0--200 MeV (II)---and
the set of cutoff radii $(R_{\rm S},R_{\rm L})$ considered---either (0.8,1.2) fm (a) or (0.7,1.0) fm (b).
In combination with each of these, the LECs $c_D$ and $c_E$ that characterize the contact terms
in the three-nucleon interaction~\cite{Piarulli:2018} have been constrained by fitting the trinucleon
binding energies and Gamow-Teller matrix element contributing to tritium $\beta$ decay~\cite{Baroni:2018}.
We note that in an earlier version of these three-nucleon interactions, rather than the Gamow-Teller
matrix element, the neutron-deuteron doublet scattering length had been reproduced~\cite{Piarulli:2018}.
The corresponding Hamiltonians, designated as NV2+3-Ia/b and NV2+3-IIa/b (or simply Ia/b and IIa/b),
will also be used below in some cases\footnote{For two-body observables,
such as the deuteron magnetic form factor, photodisintegration and threshold
electrodisintegration cross sections of interest here, the Hamiltonians Ia$^*$/b$^*$
and IIa$^*$/b$^*$ are the same as Ia/b and IIa/b, since three-nucleon interactions
are obviously not included.}.
\begin{table}[bth]
\begin{tabular}{c |c}
                         & Eqs. \\
\hline
LO &  (\ref{eq:jlo}) \\
NLO & (\ref{eq:nlor2})  \\
N2LO(RC) &(\ref{eq:j1rcc})  \\
N2LO($\Delta$) &(\ref{eq:jdn2lo}) \\
N3LO(LOOP)  &(\ref{eq:jloop})+(\ref{eq:jloopct})  \\
N3LO(MIN) &(\ref{eq:jmin}) \\
N3LO(NM)  &(\ref{eq:jnm}) \\
N3LO(OPE) & (\ref{eq:e212}) \\
\hline
\end{tabular}
\caption{Notation adopted for the various terms in the current operator; N2LO and N3LO denote
respectively the sum of all terms at N2LO and N3LO.}
\label{tb:tb1a}
\end{table}

Magnetic form factors and magnetic moments of spin $J\,$=1/2 and 1 systems can be obtained
by evaluating the matrix element~\cite{Carlson:2015}
\begin{equation}
\label{eq:ffff}
F_M(q;A)= - i\,  \frac{2\, m}{q}\, \langle A;JJ |j_y(q\, \hat{\bf x})|A; JJ\rangle \ , 
\qquad \mu_A = F_M(0;A) \ ,
\end{equation}
where $|A; JJ\rangle$ represents the ground state of the nucleus $|A; JM_J\rangle$  in the stretched configuration
having $M_J\,$=$\,J$, and $j_y(q\, \hat{\bf x})$ is the $y$-component of the current operator with
the momentum transfer ${\bf q}$ taken in the $x$-direction.  Both $A\,$=2 and 3 matrix elements
of interest here have been calculated by Monte Carlo integration techniques based on the
Metropolis algorithm~\cite{Metropolis:1953} and
utilizing random walks with, respectively, $\sim\,$$10^6$ and $\sim\,$$\,5\times 10^5$ samples
(see Ref.~\cite{Carlson:2015} for
a recent review of Monte Carlo methods as applied in nuclear physics). Statistical errors are typically
well below 1\% for each individual contribution to the current (in fact, at the level of a few parts in $10^4$ for the LO
and, typically, a few parts in $10^3$ for the higher orders),
and will not be quoted in the results reported below unless explicitly noted.
\begin{center}
\begin{table}[bth]
\begin{tabular}{l ||S|S|S|S} 
        & Ia$^*$ &  Ib$^*$ &IIa$^*$  & IIb$^*$ \\
\hline
LO                      & 0.8498 &  0.8485   & 0.8501  &   0.8501  \\
N2LO(RC)                 & -0.0062 & -0.0061 & -0.0065   & -0.0072   \\
N3LO(MIN)        & 0.0002  &  0.0005   & 0.0002   &  0.0009   \\
N3LO(NM)         &0.0093  & 0.0211    &  0.0110 & 0.0396   \\
N3LO(OPE)      &0.0042  & -0.0065   & 0.0026  & -0.0260   \\
\hline
\end{tabular}
\caption{Individual contributions to the deuteron magnetic moment in units of n.m.,
corresponding to the nuclear Hamiltonians Ia$^*$/b$^*$ and IIa$^*$/b$^*$.
The experimental value is $0.8574$ n.m., and is reproduced by adding all contributions.}
\label{tb:mud}
\end{table}
\end{center}

Individual contributions, associated with the various terms as designated in
Table~\ref{tb:tb1a}, to $\mu_d$, and $\mu_S$ and $\mu_V$,
are reported in Tables~\ref{tb:mud} and~\ref{tb:mu3}, respectively.
The NLO and N3LO(LOOP) current operators are isovector,
and therefore do not contribute to isoscalar observables, such
as $\mu_d$ (and the deuteron magnetic form factor, see below).
At N3LO the only non-vanishing contributions to isoscalar observables are those
from the OPE and minimal (MIN) and NM contact currents. 
Note, however, that since the HH trinucleon wave functions include
components with total isospin 3/2 induced by isospin-symmetry-breaking terms
from strong and electromagnetic interactions, purely isovector current
operators---specifically, those at NLO, N2LO($\Delta$), and N3LO(LOOP)---give
tiny contributions to $\mu_S$; conversely, the purely isoscalar current operator
N3LO(OPE) gives a tiny contribution to $\mu_V$.  Lastly, the sums of all
contributions in Tables~\ref{tb:mud} and~\ref{tb:mu3}
reproduce (by design) the experimental values for $\mu_d$, and $\mu_S$ and $\mu_V$.
\begin{center}
\begin{table*}[bth]
\begin{tabular}{l ||S|S|S|S||S|S|S|S}
    & \multicolumn{4}{l||}{\hspace{2.8cm}$\mu_S$}
    & \multicolumn{4}{l}{\hspace{2.8cm}$\mu_V$} \\
    \hline
       &Ia$^*$  & Ib$^*$   & IIa$^*$   & IIb$^*$ & Ia$^*$  & Ib$^*$   & IIa$^*$   & IIb$^*$ \\
\hline
LO  &   0.4089 &   0.4075 &   0.4091 &   0.4089 &  -2.1823 &  -2.1755 &  -2.1815 &  -2.1787 \\
NLO  &   0.0015 &   0.0020 &   0.0012 &   0.0018 &  -0.1967 &  -0.2257 &  -0.1967 &  -0.2255 \\
N2LO  &  -0.0062 &  -0.0043 &  -0.0052 &  -0.0071 &  -0.0388 &  -0.0657 &  -0.0395 &  -0.0617 \\
N3LO(LOOP) &   0.0004 &   0.0004 &   0.0003 &   0.0002 &  -0.0290 &  -0.0233 &  -0.0287 &  -0.0205 \\
N3LO(MIN) &   0.0001 &   0.0005 &   0.0002 &   0.0010 &   0.0035 &   0.0038 &   0.0033 &   0.0035 \\
N3LO(NM) &   0.0130 &   0.0269 &   0.0148 &   0.0488 &  -0.1098 &  -0.0668 &  -0.1100 &  -0.0704 \\
N3LO(OPE) &   0.0094 &  -0.0063 &   0.0065 &  -0.0269 &  -0.0002 &  -0.0001 &   0.0000 &   0.0002 \\
\hline
\end{tabular}
\caption{Individual contributions to the isoscalar and isovector combinations of the trinucleon
magnetic moments in units of n.m., corresponding to the nuclear Hamiltonians Ia$^*$/b$^*$
and IIa$^*$/b$^*$.  The experimental values are $0.4257$ n.n. and $-2.553$ n.m.,
respectively, and are reproduced by adding all contributions.}
\label{tb:mu3}
\end{table*}
\end{center}

As it can be surmised from the difference between models a and b in both classes I and II,
the LO contribution to the $A\,$=$\,$2 and 3 magnetic moments is very weakly
dependent on the pair of cutoff radii $(R_{\rm S},R_{\rm L})$ characterizing the two-
and three-nucleon interactions from which the $^2$H, and $^3$H and $^3$He wave
functions are derived.  In contrast the cutoff dependence is much more pronounced
in the case of the N2LO($\Delta$) and N3LO contributions, since for these the short-
and long-range regulators directly enter the correlation functions of the corresponding
transition operators. This cutoff dependence is in turn reflected in the significant variation
of the LECs $d_i^S$ and $d_1^V$ between models a and b.  The N2LO(RC) correction,
which is nominally suppressed by two powers of the expansion parameter $Q/\Lambda_\chi$,
being inversely proportional to the cube of the nucleon mass, itself of order $\Lambda_\chi$,
is in fact further suppressed than the naive N2LO power counting would imply; as a matter
of fact, it is typically an order of magnitude smaller than the N2LO($\Delta$) contribution.
\begin{figure}[bth]
\includegraphics[width=4in]{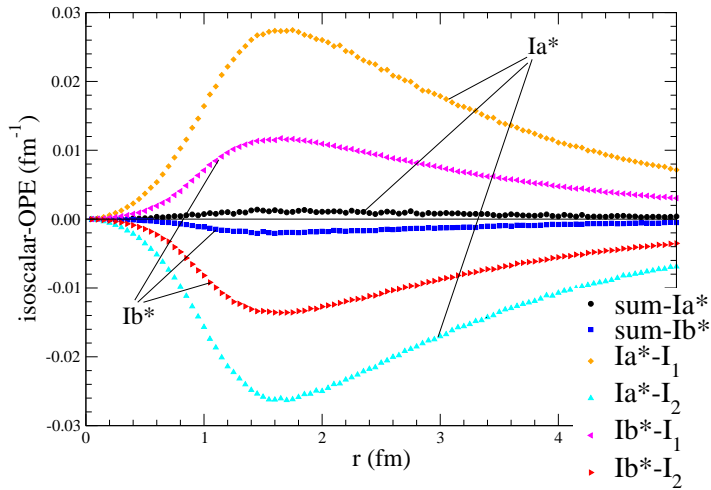}
\caption{(Color online). The ``densities'' relative to $\mu_d$ and corresponding to the terms
proportional to $I_1^\pi$ and $I_2^\pi$ in the N3LO(OPE) current of Eq.~(\ref{eq:e212})
(curves labeled $I_1$ and $I_2$) as well to the full current (curve labeled ``sum''), shown
as functions of the internucleon separation $r$ for models Ia$^*$ and Ib$^*$.
}
\label{fig:dope}
\end{figure}

The isoscalar N3LO(OPE) contribution to $\mu_d$ and $\mu_S$
exhibits the most striking cutoff dependence---it changes sign in going from models a$^*$ to b$^*$. 
The origin of this dependence becomes apparent when the contributions of the terms
proportional to the correlation functions $I_1^\pi(\mu)$ and $I_2^\pi(\mu)$ in the matrix
element of the ${\bf j}^{\rm N3LO}_{\rm OPE}$ current are calculated independently: they
turn out to be large and of opposite sign.  In Fig.~\ref{fig:dope} we show the ``densities'',
as functions of the relative distance $r$ between the two nucleons, associated with these
terms (curves labeled $I_1$ and $I_2$) as well as with their sum for the deuteron magnetic
moment matrix element---integration over $r$ gives the corresponding contribution,
in particular integration of the curve labeled ``sum'' leads to the full N3LO(OPE) contribution
to $\mu_d$ in Table~\ref{tb:mud}. The $I_1$ and $I_2$ curves are very sensitive to
the cutoff radius $R_{\rm L}$ (differences between models a and b), and for a given
model almost completely cancel each other out.

\begin{figure}[bth]
\includegraphics[width=3.75in]{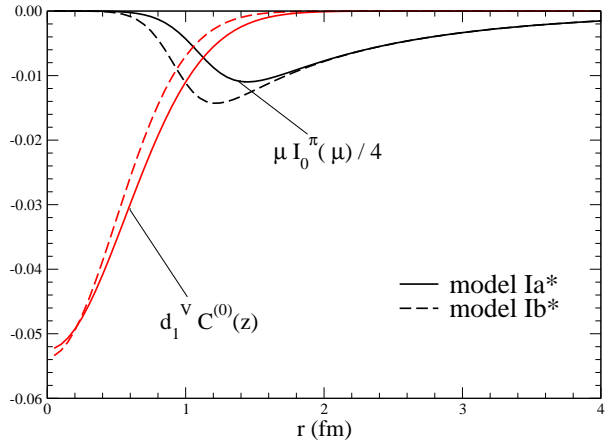}
\caption{(Color online).  Regularized correlation functions associated with the NLO and N3LO(NM) magnetic
moment operators.  }
\label{fig:ftn}
\end{figure}

The isovector contributions to $\mu_V$ at N3LO, in particular those from N3LO(NM), are not
suppressed by $(m_\pi/\Lambda_\chi)^2$ relative to the NLO ones, even though $d_1^V$
is of natural size.  However, the argument outlined above ignores the fact that the
(regularized) correlation functions of these NLO and N3LO(NM) currents have drastically
different magnitudes and ranges.
To illustrate this point, consider the isovector magnetic moment operators\footnote{The magnetic moment
operator easily follows from $-(i/2) \,{\bm \nabla}^q \times {\bf j}({\bf q}) |_{q=0}$, see Ref.~\cite{Pastore:2009} .}
\begin{eqnarray}
{\bm \mu}^{\rm NLO}&=&-\frac{1}{4\, m_\pi}  \left({\bm \tau}_i \times {\bm \tau}_j\right)_z
\, \mu_{ij}\, I_0^\pi(\mu_{ij}) \left[  {\bm \sigma}_i\times\hat{\bf r}_{ij}
 \,\, {\bm \sigma}_j \cdot \hat{\bf r}_{ij}- {\bm \sigma}_j\times\hat{\bf r}_{ij}
 \,\, {\bm \sigma}_i \cdot \hat{\bf r}_{ij}\right] + \dots \ , \\
{\bm \mu}^{\rm N3LO}({\rm NM})&=& -\frac{1}{m_\pi} (\tau_{i,z} - \tau_{j,z}) \, d_1^V\, C^{(0)}_{R_{\rm S}}(z_{ij})\,
 \left({\bm \sigma}_i-{\bm \sigma}_j\right) \ ,
 \end{eqnarray} 
 where the $\dots$ indicate additional terms from the pion-in-flight current that are ignored here for simplicity.
 The correlation functions $\mu\, I_0^\pi(\mu)/4$ and $d_1^V\, C^{(0)}(z)$ are shown
 in Fig.~\ref{fig:ftn} for the two sets of cutoff radii $(R_{\rm S},R_{\rm L})\,$=(0.8,1.2) fm for model Ia$^*$
 and (0.7,1.0) fm for model Ib$^*$.

\begin{center}
\begin{table*}[bth]
\begin{tabular}{l ||S|S||S|S}
    & \multicolumn{2}{l||}{\hspace{1cm}$\mu(^3$He)}
    & \multicolumn{2}{l}{\hspace{1cm}$\mu(^3$H)} \\
    \hline
       &LO & \hbox{N3LO}   & LO  & \hbox{N3LO}  \\
\hline
Ia  &  -1.769  & -2.119    & 2.585   &    2.969\\
Ib &  -1.765   &  -2.122  & 2.579  &  2.970    \\
IIa & -1.770  & -2.126   &  2.588    &  2.973   \\
IIb &  -1.769   &  -2.131  & 2.586   &    2.981  \\
\hline
\end{tabular}
\caption{The $^3$He and $^3$H
magnetic moments in units of n.m., corresponding to the nuclear Hamiltonians Ia/b
and IIa/b with current operators at LO and up to N3LO.  The experimental values are
$-2.126$ n.n. and $2.979$ n.m., respectively.}
\label{tb:mu3prl}
\end{table*}
\end{center}
Finally, in Table~\ref{tb:mu3prl} we report the $^3$He and $^3$H magnetic moments obtained
with the Hamiltonians Ia/b and IIa/b~\cite{Piarulli:2018} which differ from Ia$^*$/b$^*$ and
IIa$^*$/b$^*$ only in the values adopted for the LECs $c_D$ and $c_E$ in the three-nucleon contact
interaction.  The results obtained with the three LECs in Table~\ref{tb:tb1} (that is, without refitting
$\mu_d$, $\mu_S$ and $\mu_V$) and by summing all corrections
up to N3LO are well within less than a \% of the experimental values, indicating that the wave
functions of models a and b are close to those of models a$^*$ and b$^*$ (which reproduce these
experimental values by design).  As shown below, this conclusion remains valid also in the case
of the magnetic form factors for momentum transfers $\lesssim 3$ fm$^{-1}$.
\section{Predictions for selected electromagnetic observables}
\label{sec:res}
In this section we present predictions for the magnetic form factors of $^2$H, $^3$He,
and $^3$H, deuteron photodisintegration at low energies (up to 30 MeV), and deuteron
threshold electrodisintegration at backward angles up to four-momentum transfers
$Q^2\lesssim 30$ fm$^{-2}$. (In this section, $Q$ denotes, rather than the generic low-momentum
scale introduced earlier, the four-momentum transfer defined as $Q^2\,$=$\,q^2-\omega^2$,
where $q$ and $\omega$ are the three-momentum and energy transfers, respectively.)
There are extensive sets of experimental data on elastic
electromagnetic cross sections and polarization observables of few-nucleon systems, and
an up-to-date list of references to these can be found in the recent compilation by Marcucci
{\it et al.}~\cite{Marcucci:2016}.  The experimental values for the form factors presented in the
figures below result from fits to these data sets (the {\it world data})---see Ref.~\cite{Marcucci:2016}
for a discussion of the procedure utilized to carry out the analysis.  Experimental data on the deuteron low-energy
photodisintegration and threshold electrodisintegration cross sections are from, respectively,
Refs.~\cite{Bishop:1950,Snell:1950,Colgate:1951,Carver:1951,Birenbaum:1985,Moreh:1989,DeGraeve:1992}
and~\cite{Rand:1967,Ganichot:1972,Simon:1979,Bernheim:1981,Auffret:1985,Schmitt:1997} (the
deuteron electrodisintegration  measurements at SLAC~\cite{Arnold:1990,Frodyma:1993} are not considered here,
since for these $Q^2 \gtrsim 30$ fm$^{-2}$).  The electrodisintegration data have been averaged
over the interval 0--3 MeV of the recoiling $np$ center-of-mass energy (note that for the
SLAC data this interval was 0--10 MeV).

Before comparisons with experimental data can be made, however, we need
to include hadronic electromagnetic form factors in the current
operators of Sec.~\ref{sec:em3}. These could be consistently
calculated in chiral perturbation theory~\cite{Kubis:2001}, but the convergence of these calculations in
powers of the momentum transfer appears to be rather poor.  For this reason, in the results
reported below for the $A\,$=$\,$2--3 form factors and deuteron electrodisintegration, they are taken from
fits to available electron scattering data, as detailed in Ref.~\cite{Piarulli:2013}; specifically,
in the LO and N2LO(RC) currents the replacements 
\begin{eqnarray}
\epsilon_k  \longrightarrow  \left[ G_E^S(Q^2)+G_E^V(Q^2)\, \tau_{k,z}\right]\!/2 \qquad {\rm and}\qquad
 \mu_k \longrightarrow \left[ G_M^S(Q^2)+G_M^V(Q^2)\, \tau_{k,z}\right]\!/2  \ ,
\label{eq:ekmn}
\end{eqnarray}
are made, where $G^{S/V}_E(Q^2)$ and $G^{S/V}_M(Q^2)$ denote the isoscalar/isovector
combinations of the proton and neutron electric ($E$) and magnetic ($M$)
form factors, normalized as $G^S_E(0)\,$=$\,G^V_E(0)$=$\,1$, $G^S_M(0)\,$=$\,\mu^S$,
and $G^V_M(0)\,$=$\,\mu^V$ (we use the dipole parameterization~\cite{Marcucci:2016}, including the
Galster factor for the neutron electric form factor). The NLO and N3LO(LOOP) currents, and the isovector (isoscalar)
terms of the N3LO(MIN) and N3LO(NM) currents, are multiplied by $G_E^V(Q^2)$ [$G_E^S(Q^2)$].
While for the NLO, N3LO(LOOP), and N3LO(MIN) currents a reasonable
argument can be made based on current conservation for multiplying them by
$G_E^S(Q^2)$ and $G_E^V(Q^2)$~\cite{Piarulli:2013,Marcucci:2016}, there is no {\it a priori}
justification for the use of these nucleon form factors in the non-minimal contact current---they are simply
included in order to provide a reasonable falloff of the interaction vertex
with increasing $Q^2$.  Lastly, the N2LO($\Delta$) current is multiplied
by the $\gamma N\Delta$ electromagnetic form factor, taken from an analysis of $\gamma N$ data
in the $\Delta$-resonance region~\cite{Carlson:1986} and parametrized as
\begin{equation}
\frac{G_{\gamma N \Delta}(Q^2)}{\mu_{ \Delta N} }= \frac{1}
{( 1+Q^2/\Lambda_{\Delta,1}^2 )^2
\sqrt{1+Q^2/\Lambda_{\Delta,2}^2} } \ ,
\end{equation}
where $\mu_{\Delta N}$ is the nucleon-to-$\Delta$ transition magnetic
moment introduced earlier, and cutoffs $\Lambda_{\Delta,1}\,$=$\,0.84$ GeV and
$\Lambda_{\Delta,2}\,$=$\,1.2$ GeV, while the isoscalar N3LO(OPE) current,
which in a resonance saturation picture reduces to the
$\gamma\pi\rho$ current~\cite{Pastore:2009}, is multiplied by
a $\gamma\pi\rho$ transition form factor.  By
assuming vector-meson dominance, we parametrize it as
\begin{equation}
G_{\gamma\pi\rho}(Q^2)=\frac{1}{1+Q^2/m_\omega^2} \ ,
\end{equation}
where $m_\omega$ is the $\omega$-meson mass.

\subsection{Deuteron magnetic form factor}
\begin{figure}[bth]
\includegraphics[width=3.65in]{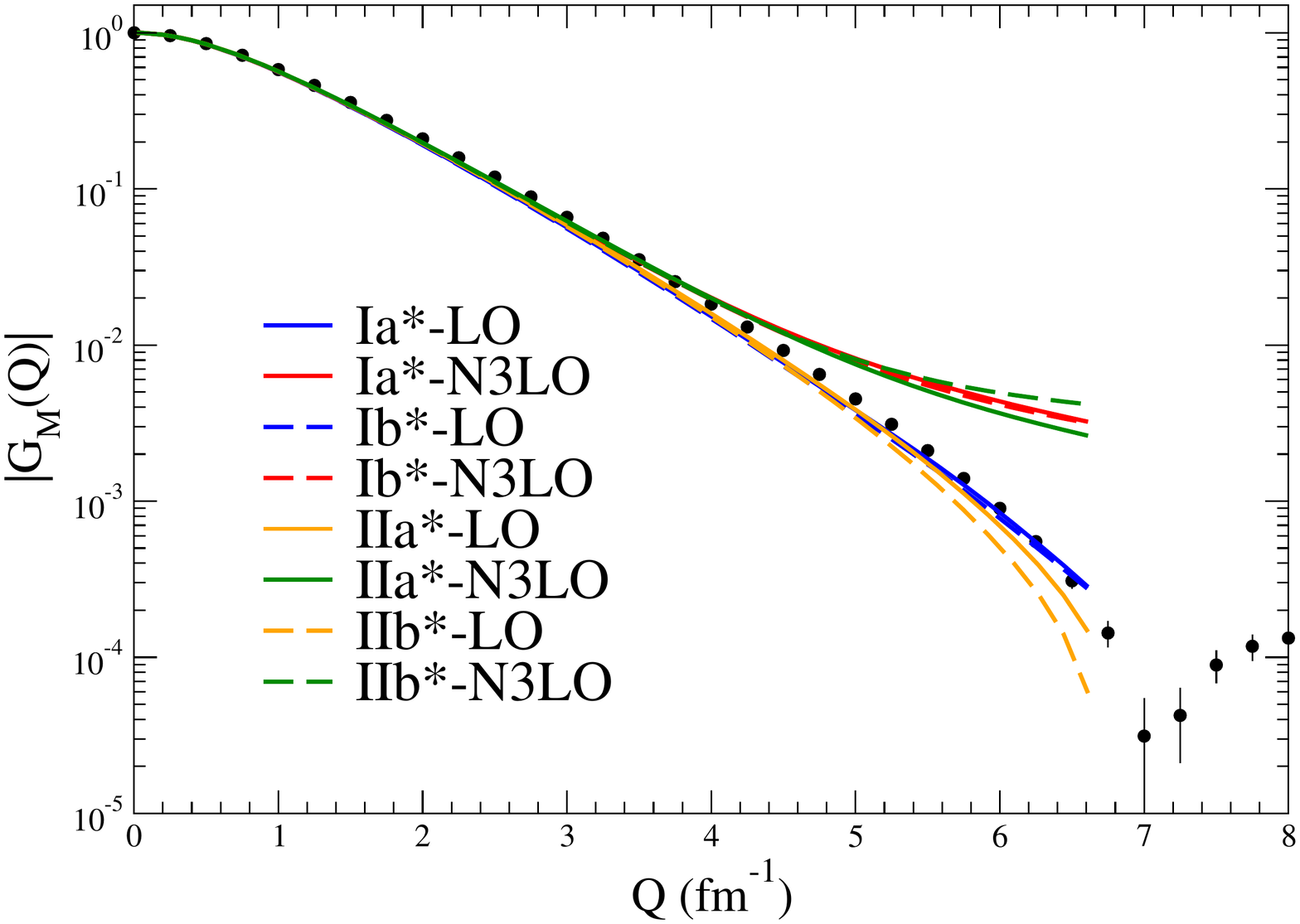}\includegraphics[width=3.65in]{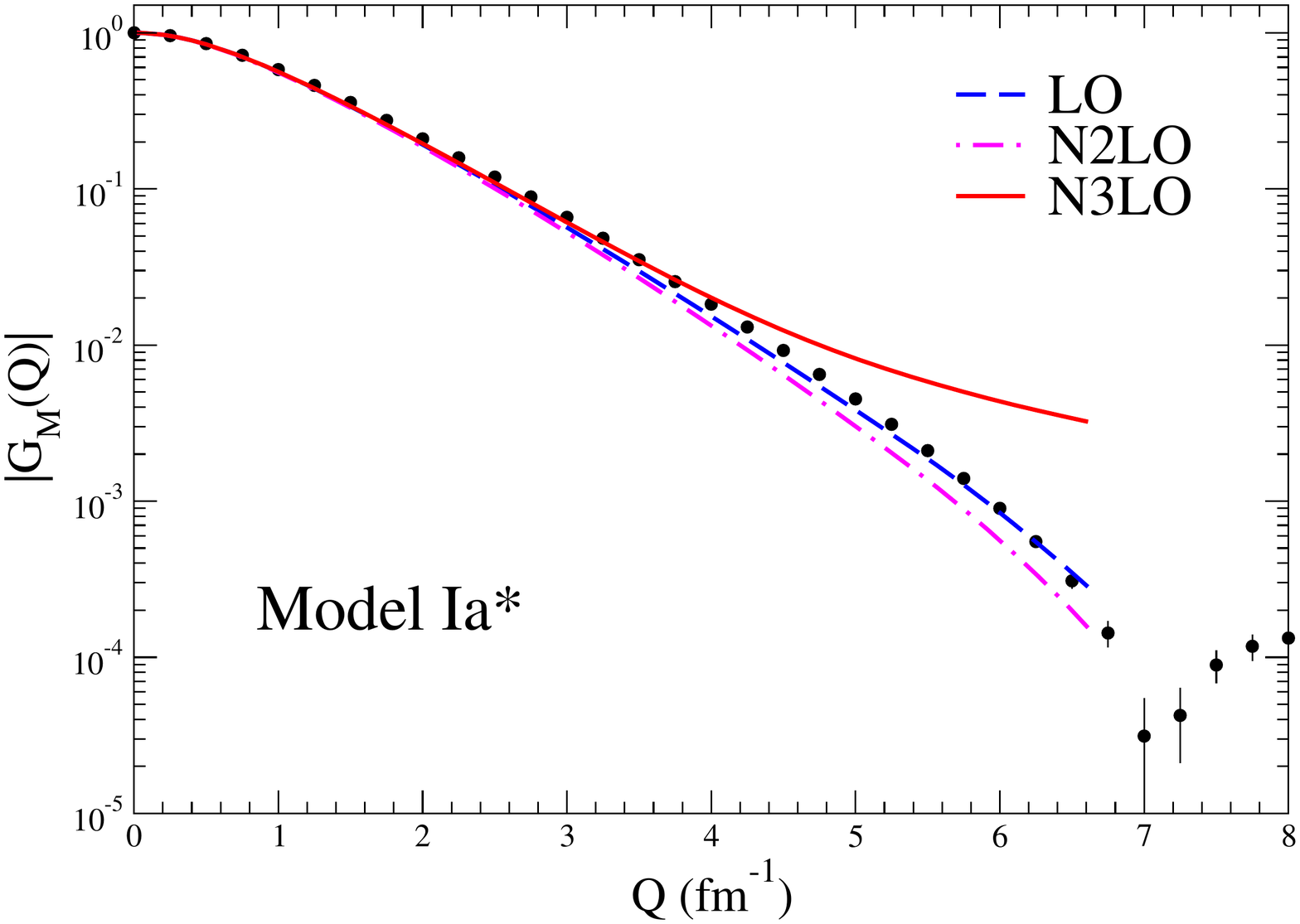}
\caption{(Color online).  Left panel: Predictions for the deuteron magnetic form factor, obtained
with currents at LO and up to N3LO for models Ia$^*$/b$^*$ and IIa$^*$/b$^*$, are compared
to the experimental data.  Right panel: Cumulative contributions to the deuteron magnetic form
factor, obtained at LO, N2LO, and N3LO for models Ia$^*$, are compared to experimental data;
note that the contributions at NLO, being isovector, vanish for this observable. }
\label{fig:fmd}
\end{figure}

The magnetic form factor obtained with models Ia$^*$/b$^*$ and IIa$^*$/b$^*$ and
currents at LO and by including corrections up to N3LO are compared to data in the
left panel of Fig.~\ref{fig:fmd}.  There is generally good agreement between theory
and experiment for four-momentum transfer values $Q$ up to $\simeq\,$3 fm$^{-1}$.
At higher $Q$'s, theory overestimates the data by a large factor, when the current
retains the N3LO corrections; in particular, the diffraction seen in the data at
$Q\,$$\simeq\,$ 7 fm$^{-1}$ is absent in the calculations.  The cutoff dependence,
as reflected by differences in the Ia$^*$ and Ib$^*$, IIa$^*$ and IIb$^*$ results, is negligible
at the lower $Q$'s and moderate at the highest $Q$'s. 

\begin{figure}[bth]
\includegraphics[width=4in]{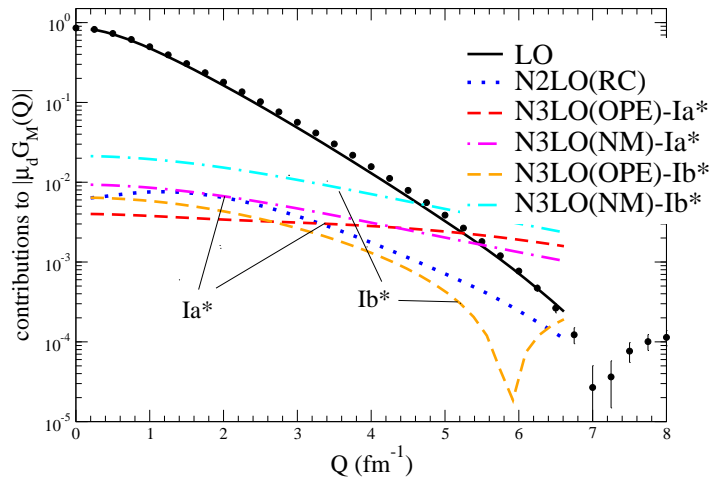}
\caption{(Color online). 
Magnitudes of individual contributions to the deuteron magnetic form factor (normalized to $\mu_d$ at $Q\,$=$\,0$) obtained with the
N3LO(OPE) and N3LO(NM) currents for models Ia$^*$ and Ib$^*$.  Note that the N3LO(MIN)
contribution is not shown, since it is much smaller than the N3LO(NM)
one (see text); also, to reduce clutter, since the LO and N2LO(RC) contributions for models Ia$^*$
and Ib$^*$ are very close to each other, they are only shown for model Ia$^*$.  The signs are as
follows: positive for LO, N3LO(OPE)-Ia$^*$, N3LO(NM) and negative for N2LO and N3LO(OPE)-Ib$^*$
(for $Q\lesssim 6$ fm$^{-1}$).}
\label{fig:fmdctr}
\end{figure}

The cumulative contributions obtained with the LO, N2LO, and N3LO currents are
illustrated for model I$a^*$ in the right panel of Fig.~\ref{fig:fmd}.  Note that only
the N2LO(RC), and  N3LO(MIN), N3LO(NM), and N3LO(OPE) currents contribute to
isoscalar observables.  In particular, the N3LO(MIN) and N3LO(NM) ones have the
same (isoscalar) operator structure and only differ in the LEC which multiplies it,
either $m_\pi^4\, C_5/8$ for N3LO(MIN) or  $d_1^S\,$=$\,m_\pi^4\, C_{15}^\prime$
for N3LO(NM).  The combination $m_\pi^4\, C_5/8$ is~\cite{Piarulli:2016}
$\simeq$ --0.000195  (--0.000199) for model Ia$^*$ (IIa$^*$) and
--0.000560 (--0.00108) for model Ib$^*$ (IIb$^*$), and should be compared to
the values for $d_1^S$ reported in Table~\ref{tb:tb1}.  Thus, the N3LO(MIN)
contribution has the same sign as, but is suppressed by more than an order of magnitude
relative to, the N3LO(NM) one.  The N3LO(NM) and N3LO(OPE) contributions have
the same (opposite) sign over the whole range of momentum transfers for models a$^*$
(b$^*$), see Fig.~\ref{fig:fmdctr}.  The N3LO(NM) for models
Ia$^*$ and Ib$^*$, and N3LO(OPE) for model Ia$^*$, are the dominant contributions
to the form factor in the high-$Q$ region.  With the choice made for hadronic
electromagnetic form factors noted above, it turns out that the momentum transfer falloff of the
N3LO(NM) contribution is simply given by the isoscalar nucleon form factor $G_E^S(Q^2)$,
since the matrix element $\langle d| j^{\rm N3LO}_{{\rm NM},y}(q\hat{\bf x})|d\rangle / q$ is
independent of $q$, see Eqs.~(\ref{eq:jnm}) and~(\ref{eq:ffff}).  The sign and magnitude of
the N3LO(OPE) contribution depend crucially on the interaction model---positive sign for models a and
negative one for models b---for the reason explained in Sec.~\ref{sec:fit}.

It is interesting to compare the results obtained here for the deuteron 
magnetic form factor with those of Refs.~\cite{Schiavilla:2002} and~\cite{Piarulli:2013}.
In Ref.~\cite{Schiavilla:2002} a calculation of $G_M(Q)$ was carried out in the conventional 
meson-exchange picture based on the Argonne $v_{18}$ (AV18) interaction~\cite{Wiringa:1995}.
In Ref.~\cite{Piarulli:2013}, instead, $G_M(Q)$ was studied both within a hybrid $\chi$EFT
approach (i.e., based on the AV18 interaction and $\chi$EFT currents), and
within a consistent $\chi$EFT approach, using the N3LO chiral interaction of Refs.~\cite{Entem:2003,Machleidt:2011}
(see also Ref.~\cite{Marcucci:2016}). In all theses cases, the complete calculation is unable to
predict the magnetic form factor in the diffraction region $Q\simeq 7$ fm$^{-1}$, either
underpredicting (in the case of the N3LO interaction) or
overpredicting (in the case of the AV18 interaction, both within the conventional and
hybrid approach) the experimental data.

\subsection{Trinucleon magnetic form factors}
\begin{figure}[bth]
\includegraphics[width=3.65in]{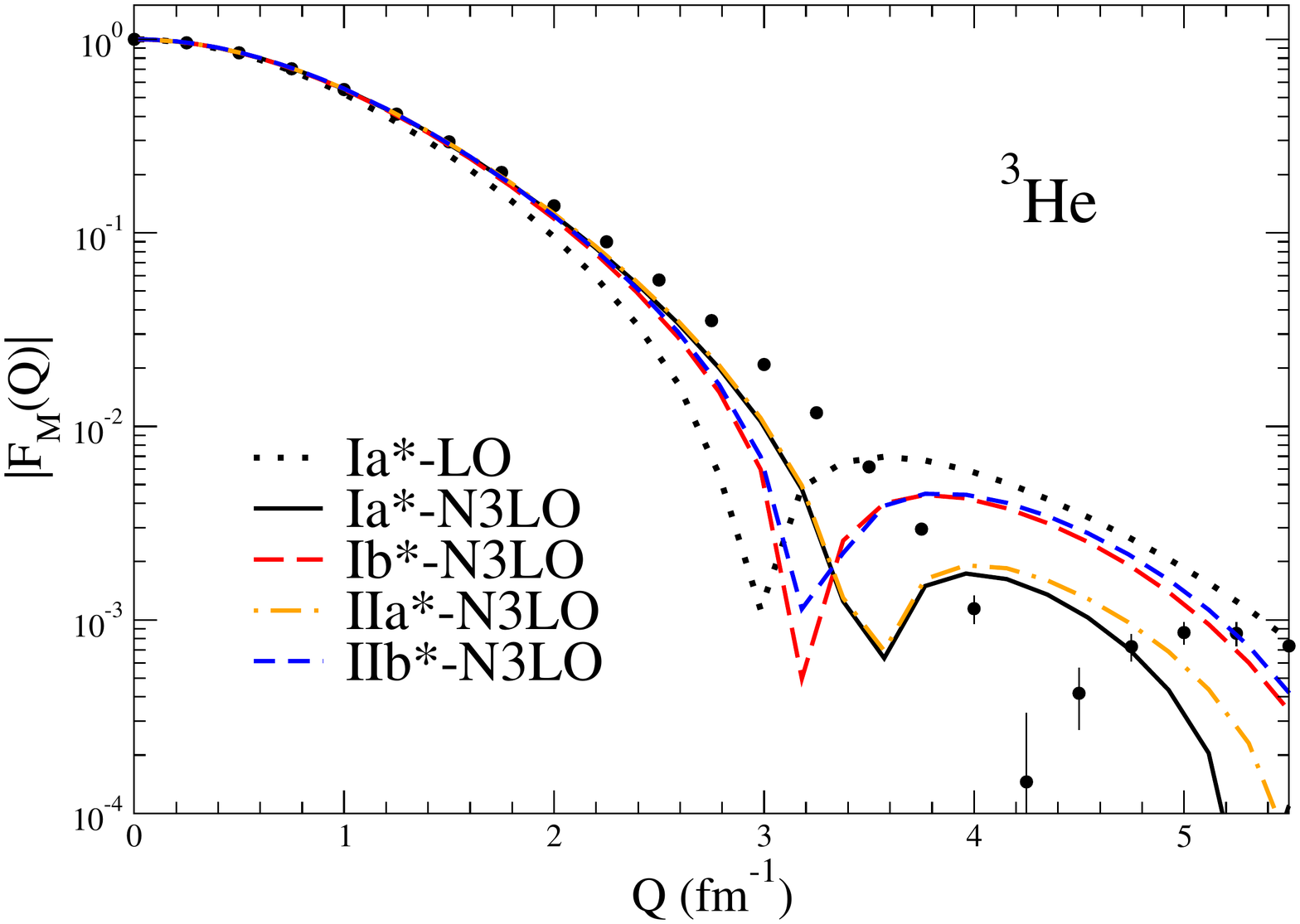}\includegraphics[width=3.65in]{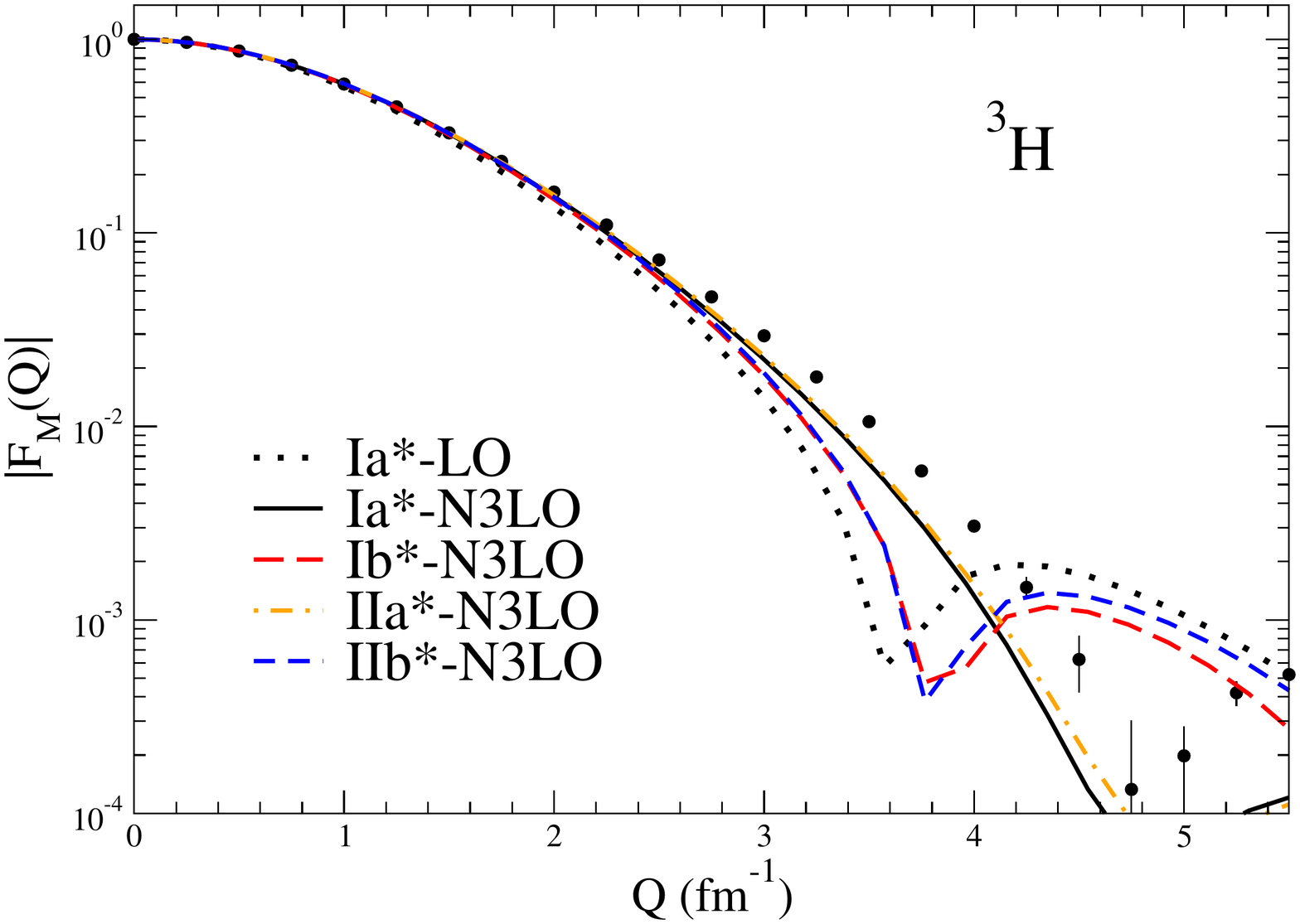}\\
\includegraphics[width=3.65in]{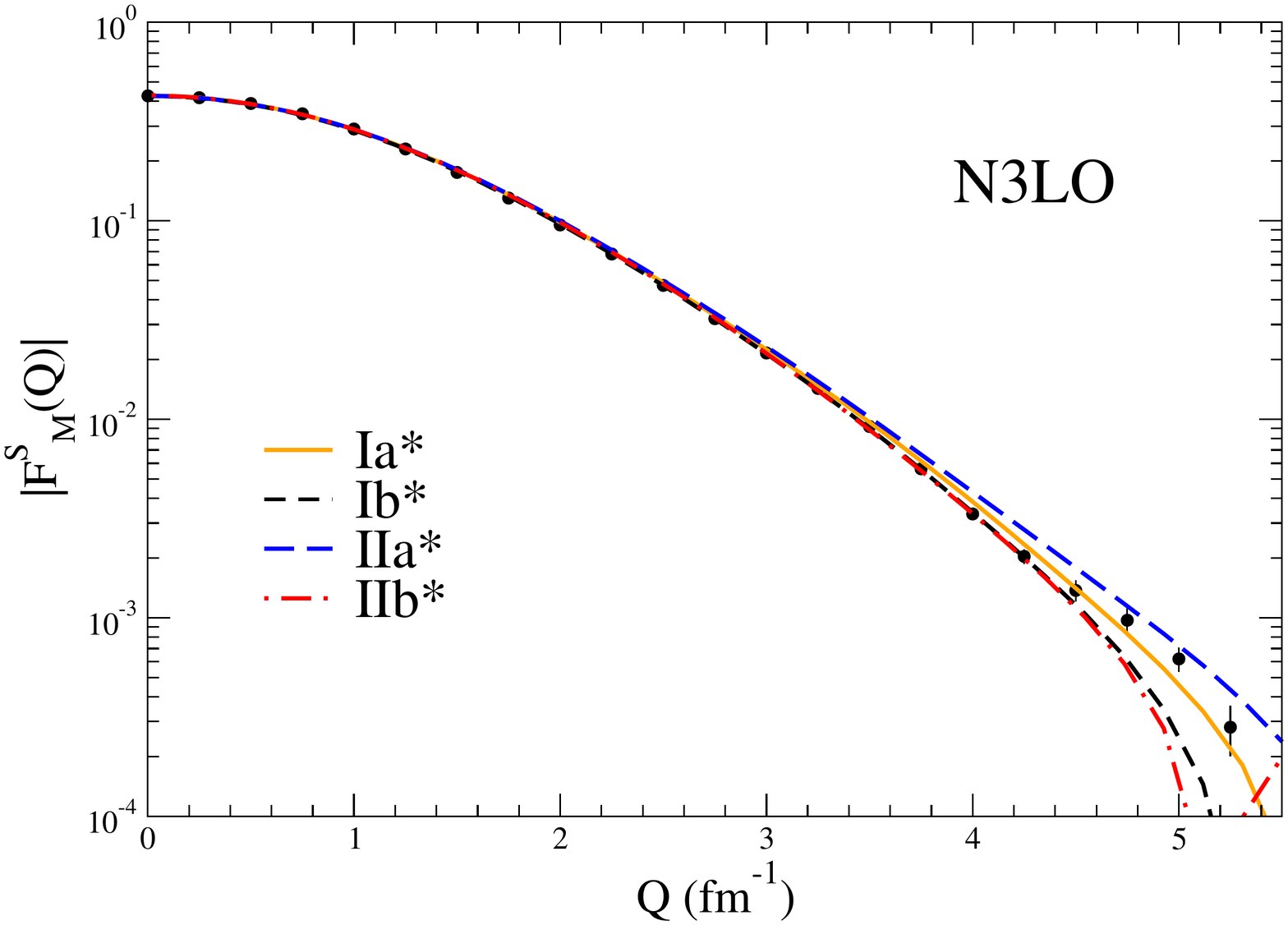}\includegraphics[width=3.65in]{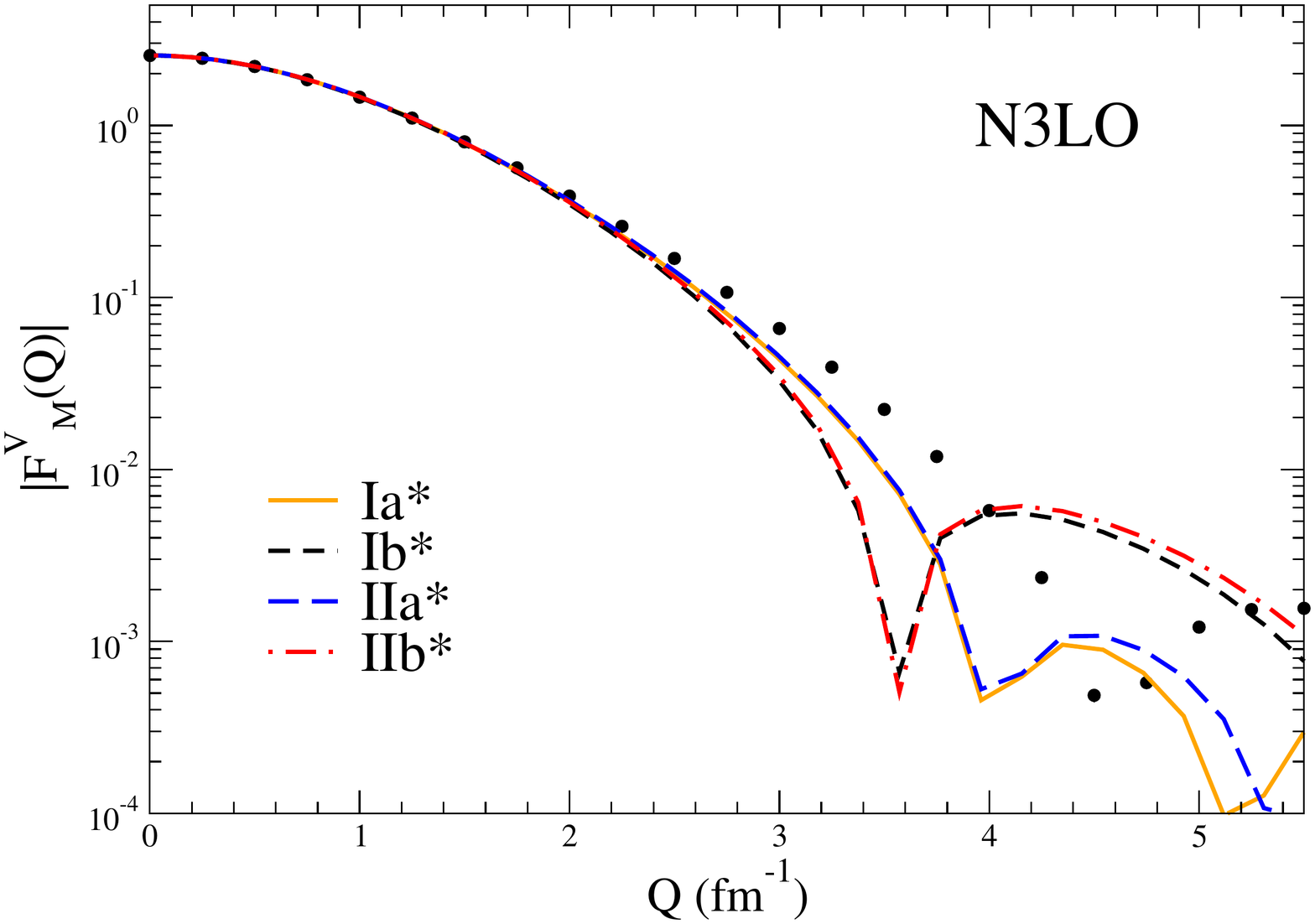}
\caption{(Color online). Top panels: Predictions for the $^3$He and $^3$H magnetic form factors obtained
with models Ia$^*$/b$^*$ and IIa$^*$/b$^*$ by including contributions up to N3LO in the current operator,
are compared to data; also shown are the results for model Ia$^*$ with the LO current operator.  Bottom
panels: Predictions for the isoscalar and isovector combinations of trinucleon magnetic form factors
obtained with models Ia$^*$/b$^*$ and IIa$^*$/b$^*$ by including contributions up to N3LO in the
current operator, are compared to data.}
\label{fig:fm3he}
\end{figure}

The magnetic form factors of $^3$He and $^3$H and their isoscalar and isovector
combinations $F_M^S(Q)$ and $F_M^V(Q)$, normalized respectively as $\mu_S$
and $\mu_V$ at $Q\,$=$\,$0, at LO for model Ia$^*$, and with inclusion of corrections up
to N3LO for all model interactions, are displayed in Fig.~\ref{fig:fm3he}.  As is well
known from studies based on the conventional meson-exchange approach (see
Ref.~\cite{Carlson:1998} and references therein), two-body current contributions are
crucial for {\it filling in} the zeros obtained in the LO calculation due to the interference
between the S- and D-state components in the ground states of the trinucleons.
For $Q\,$$\lesssim\,$2 fm$^{-1}$ there is excellent agreement between the
present predictions and experimental data.  However, as the momentum
transfer increases, even after making allowance for the significant cutoff dependence
(differences between a$^*$ and b$^*$ models), theory tends to underestimate
the data; in particular, it predicts the zeros in the $^3$He and $^3$H magnetic form factors
occurring at significantly lower values of $Q$ than observed.  Inspection of the
lower panels of Fig.~\ref{fig:fm3he} makes it clear that these discrepancies are
primarily in the isovector form factor.  Thus, the first diffraction
region remains problematic for the present models, confirming results of
calculations based both on meson-exchange phenomenology~\cite{Carlson:1998}
and earlier models of (momentum-space) chiral interactions~\cite{Entem:2003,Machleidt:2011}
and currents~\cite{Piarulli:2013}.  

\begin{figure}[bth]
\includegraphics[width=3.65in]{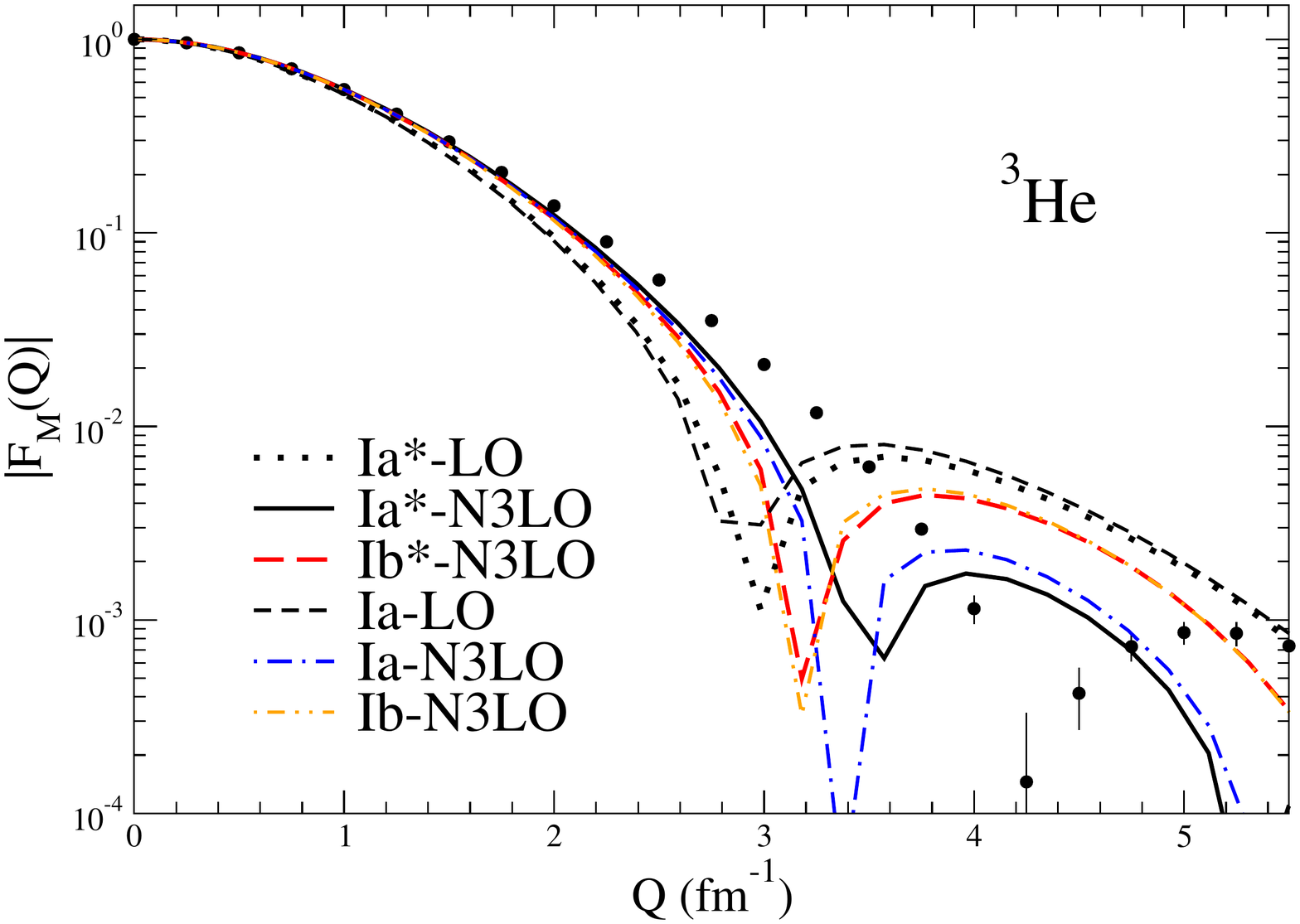}\includegraphics[width=3.65in]{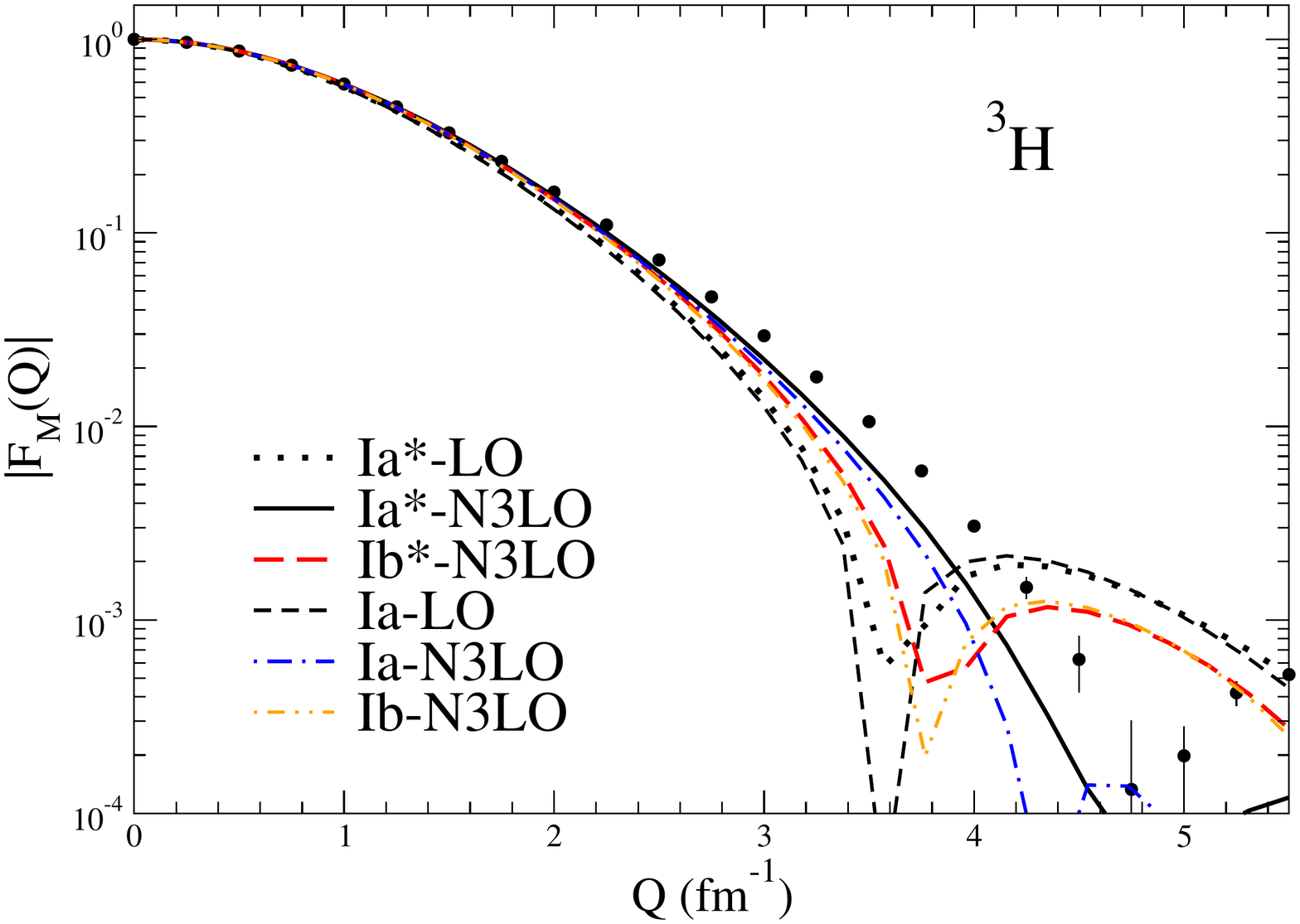}\\
\includegraphics[width=3.65in]{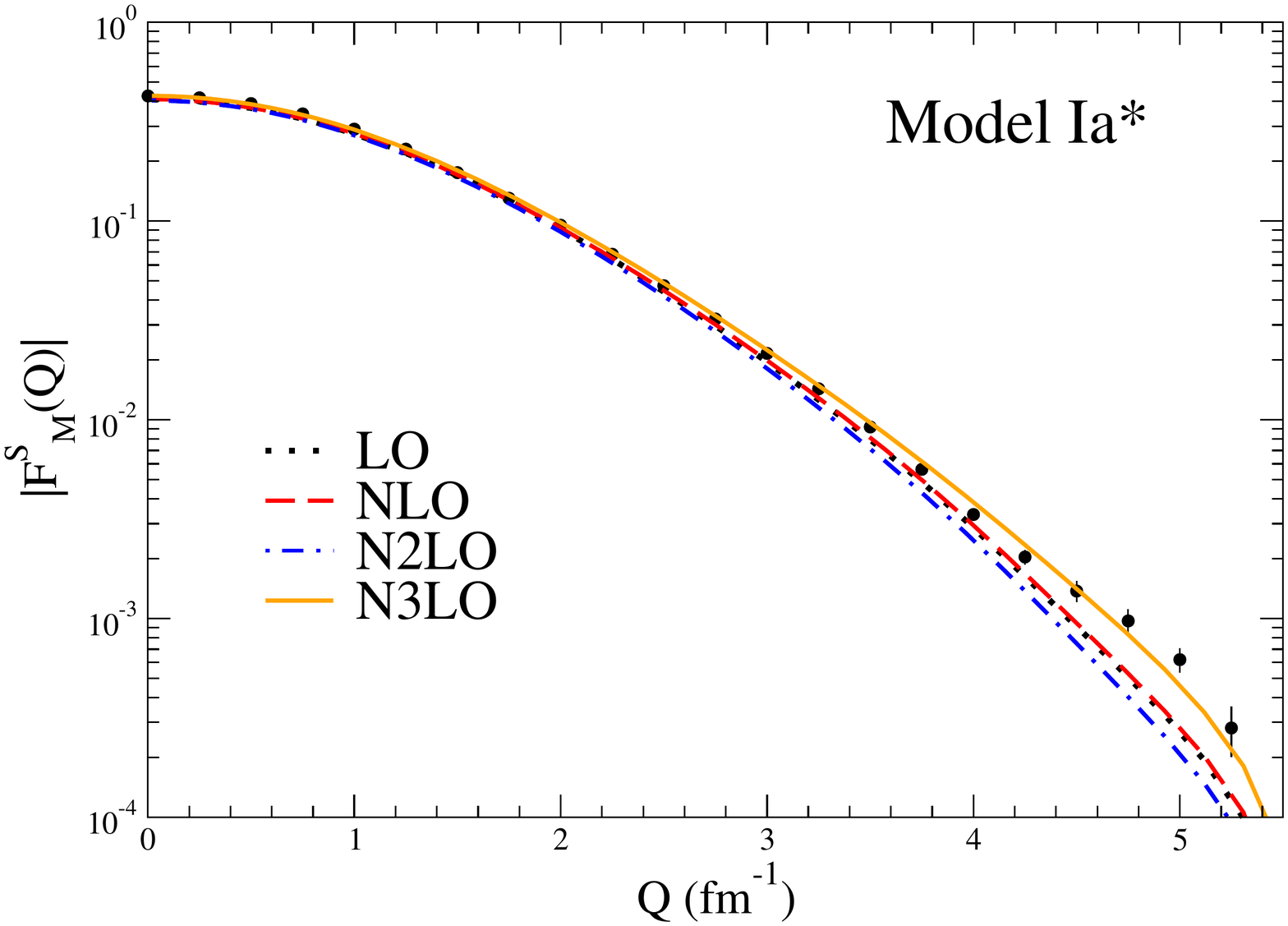}\includegraphics[width=3.65in]{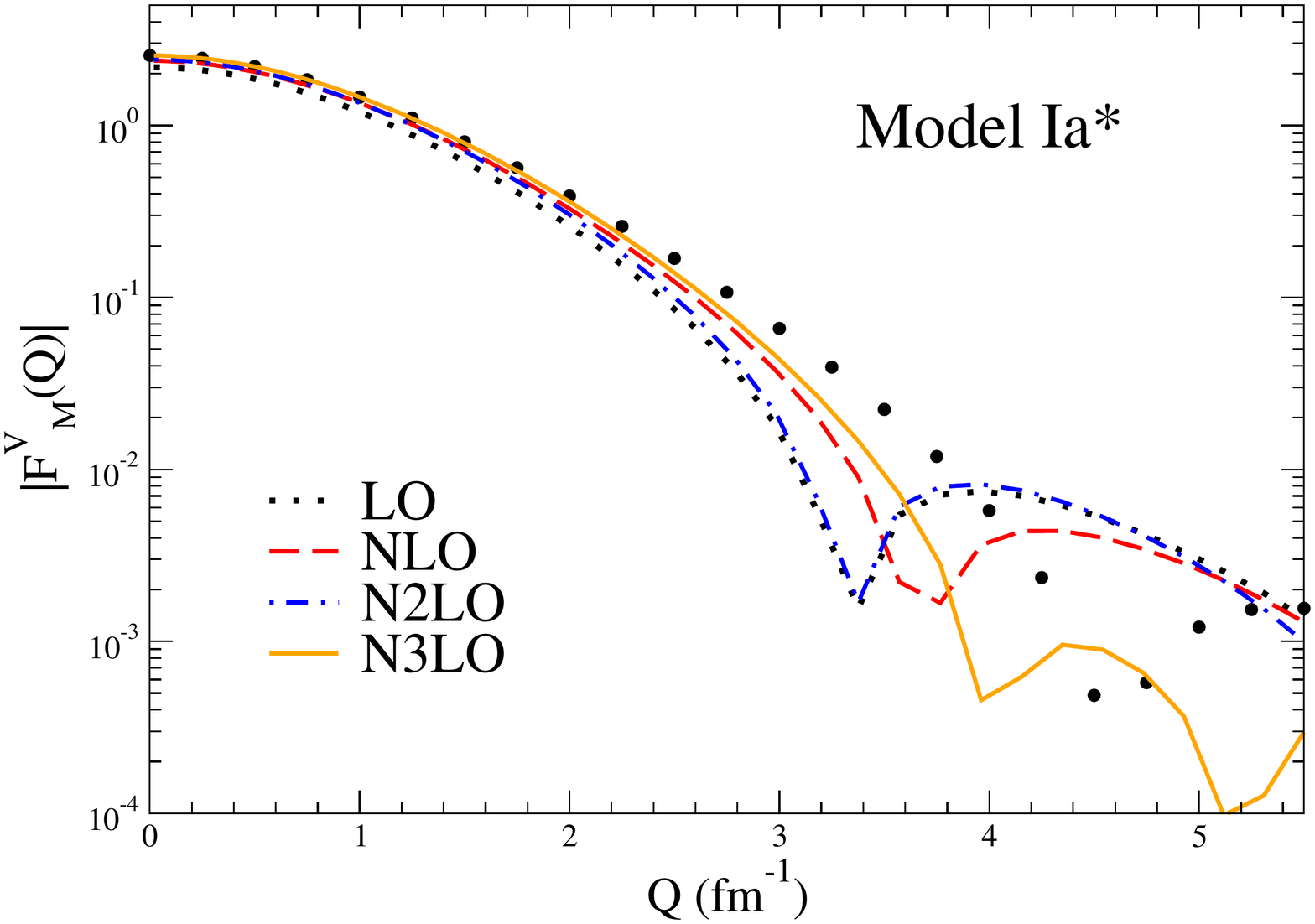}
\caption{(Color online).  Top panels: Predictions for the $^3$He
and $^3$H magnetic form factors obtained with models Ia$^*$/b$^*$ and Ia/b by including contributions
up to N3LO in the current operator, are compared to data; also shown are the results for models
Ia$^*$ and Ia with the LO current operator. Bottom panels: Cumulative contributions to the isoscalar
and isovector combinations of trinucleon magnetic form factors, obtained at LO, NLO, N2LO, and N3LO
for models Ia$^*$, are compared to the experimental data.  }
\label{fig:fmisc}
\end{figure}
\begin{figure}[bth]
\includegraphics[width=3.65in]{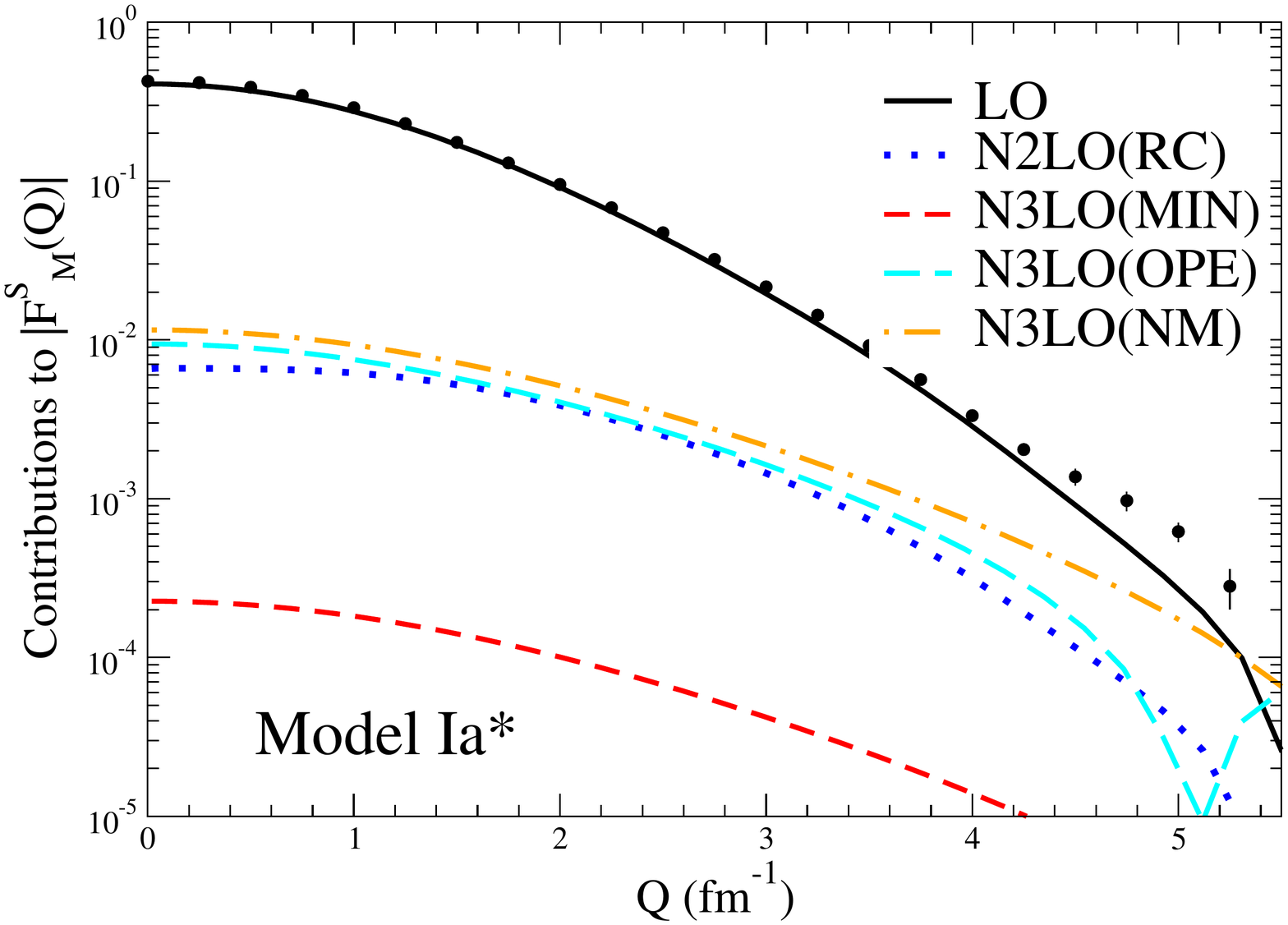}\includegraphics[width=3.65in]{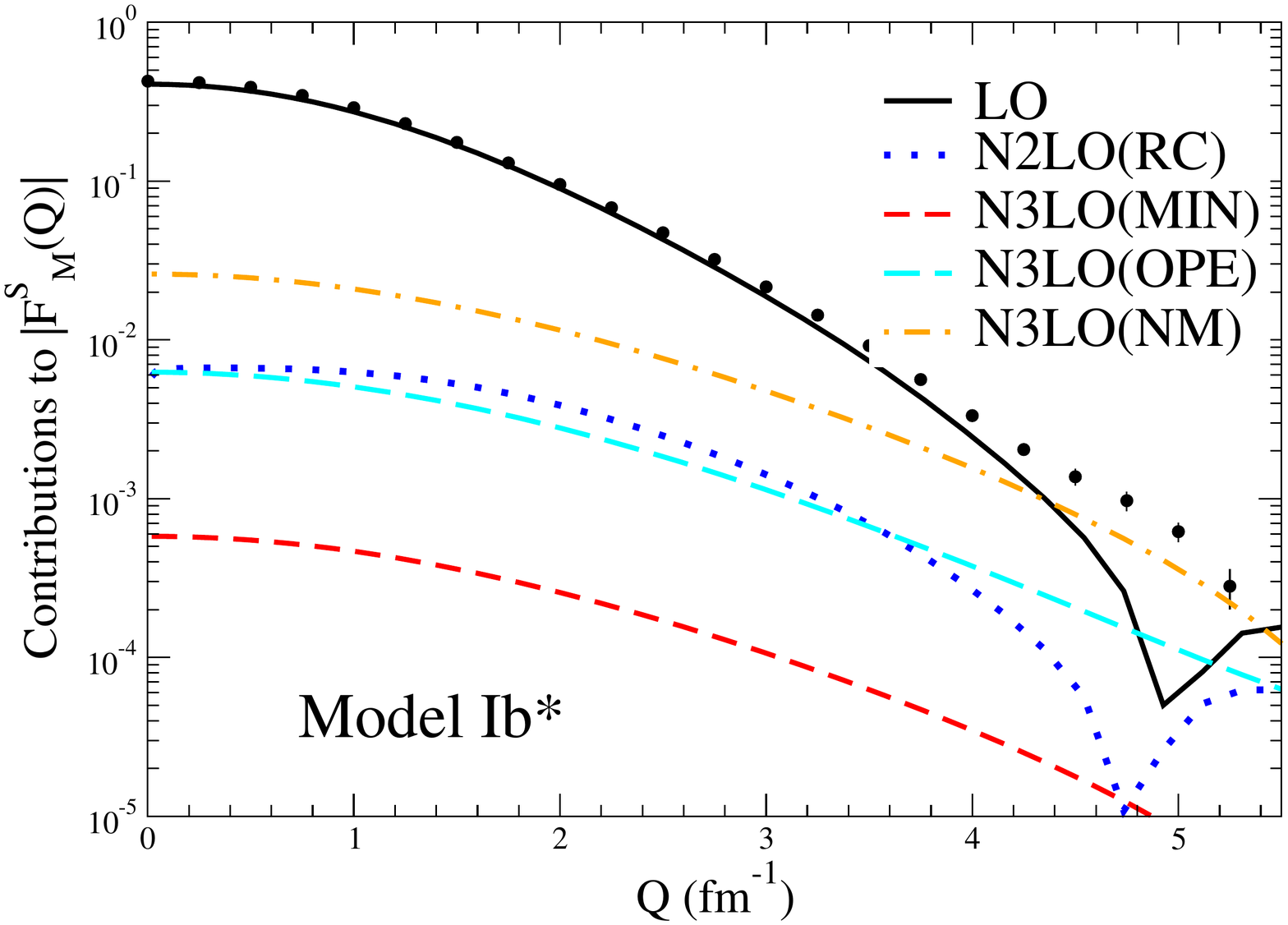}\\
\includegraphics[width=3.65in]{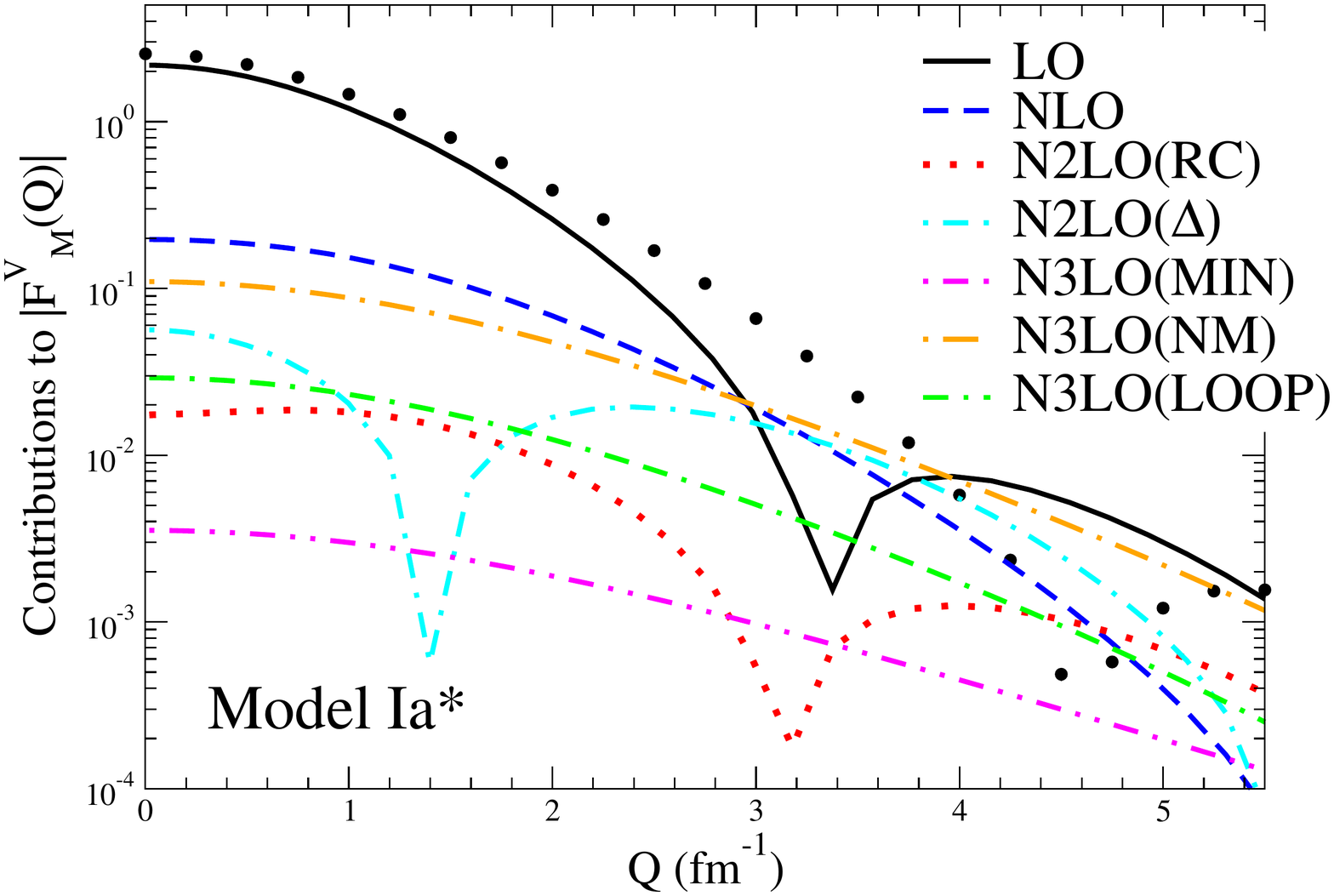}\includegraphics[width=3.65in]{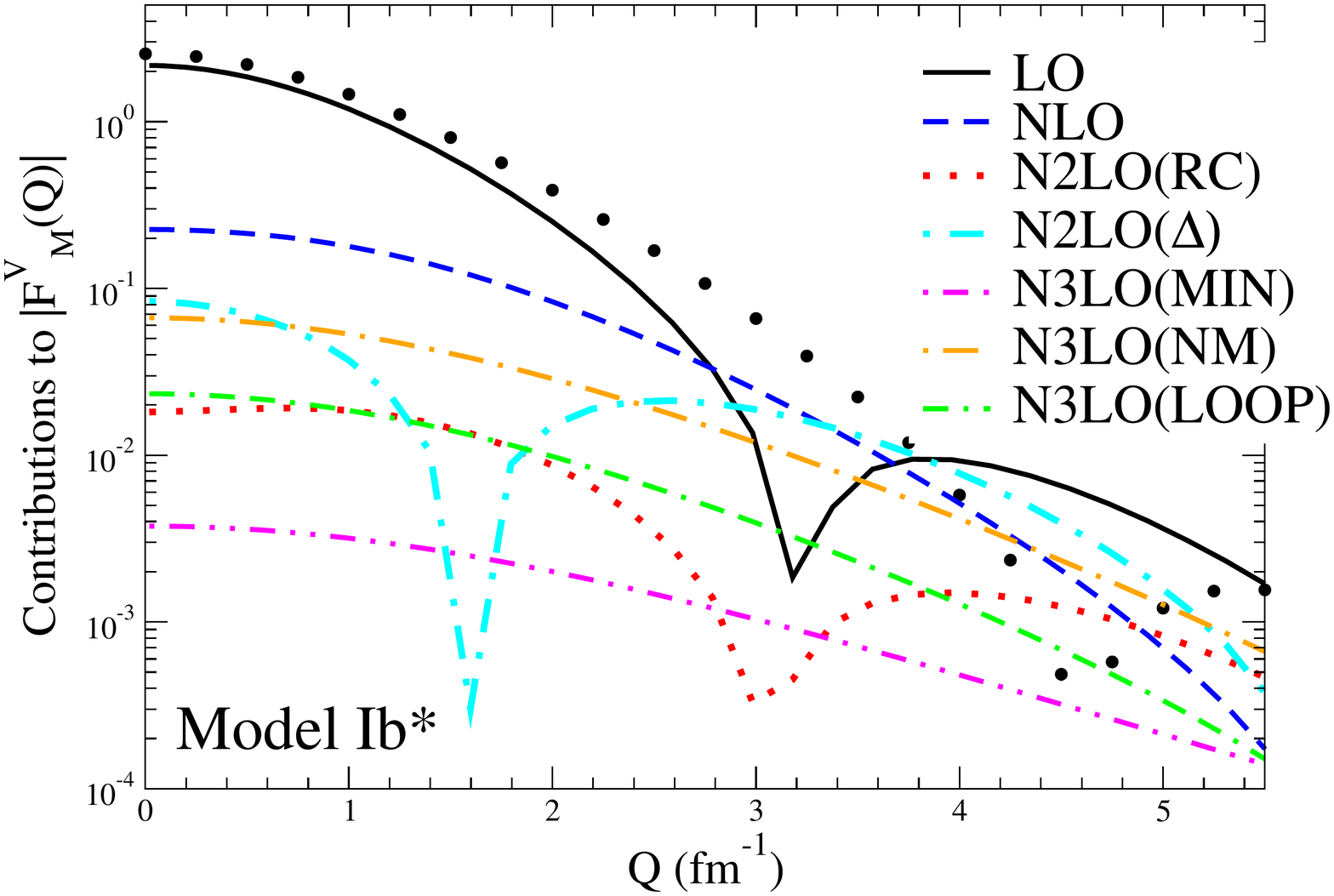}
\caption{(Color online). Magnitudes of individual contributions to the isoscalar (top panels)
and isovector (lower panels) form factors obtained for models Ia$^*$ and Ib$^*$.  The signs for $F_M^S(Q)$ are as
follows: positive for LO, N3LO(MIN), N3LO(OPE)-Ia$^*$, N3LO(NM) and negative for N2LO and N3LO(OPE)-Ib$^*$;
the signs for $F_M^V(Q)$ are as follows: negative for LO, NLO, N2LO($\Delta$) (for $Q\,$$\lesssim\,$1.5 fm$^{-1}$), N3LO(NM), and N3LO(LOOP)
and positive for N2LO(RC) and N3LO(MIN) for both models.}
\label{fig:fmcbt}
\end{figure}

The top panels of Fig.~\ref{fig:fmisc} illustrate the dependence of the LO and N3LO predictions
on the three-nucleon interaction associated with the Hamiltonian models Ia/b and Ia$^*$/b$^*$.
The variations are negligible at lower momentum transfer, and become perceptible only
in the diffraction region of the form factors.  The lower panels exhibit cumulatively the LO, NLO,
N2LO, and N3LO contributions to the isoscalar and isovector form factors, obtained with model
Ia$^*$.   Lastly, in Fig.~\ref{fig:fmcbt} the contributions of the various terms in the
current to these form factors are shown individually for models Ia$^*$ and Ib$^*$.
The sign change of the N2LO($\Delta$) contribution at $Q\,$$\simeq\,$ 1.5 fm$^{-1}$
should be noted.  As a consequence, while this contribution has the same sign
as the NLO(OPE) and N3LO(NM) ones at low $Q$ ($\lesssim\,$ 1.5 fm$^{-1}$), it interferes destructively
with them in the region $Q\,$$\simeq\,$ (3.0--3.4) fm$^{-1}$, where the LO form factor
has a zero.  This interference is largely responsible for the failure to reproduce the observed
isovector form factor in the diffraction region.  

\subsection{Deuteron photodisintegration at low energies}
\begin{figure}[bth]
\includegraphics[width=3.75in]{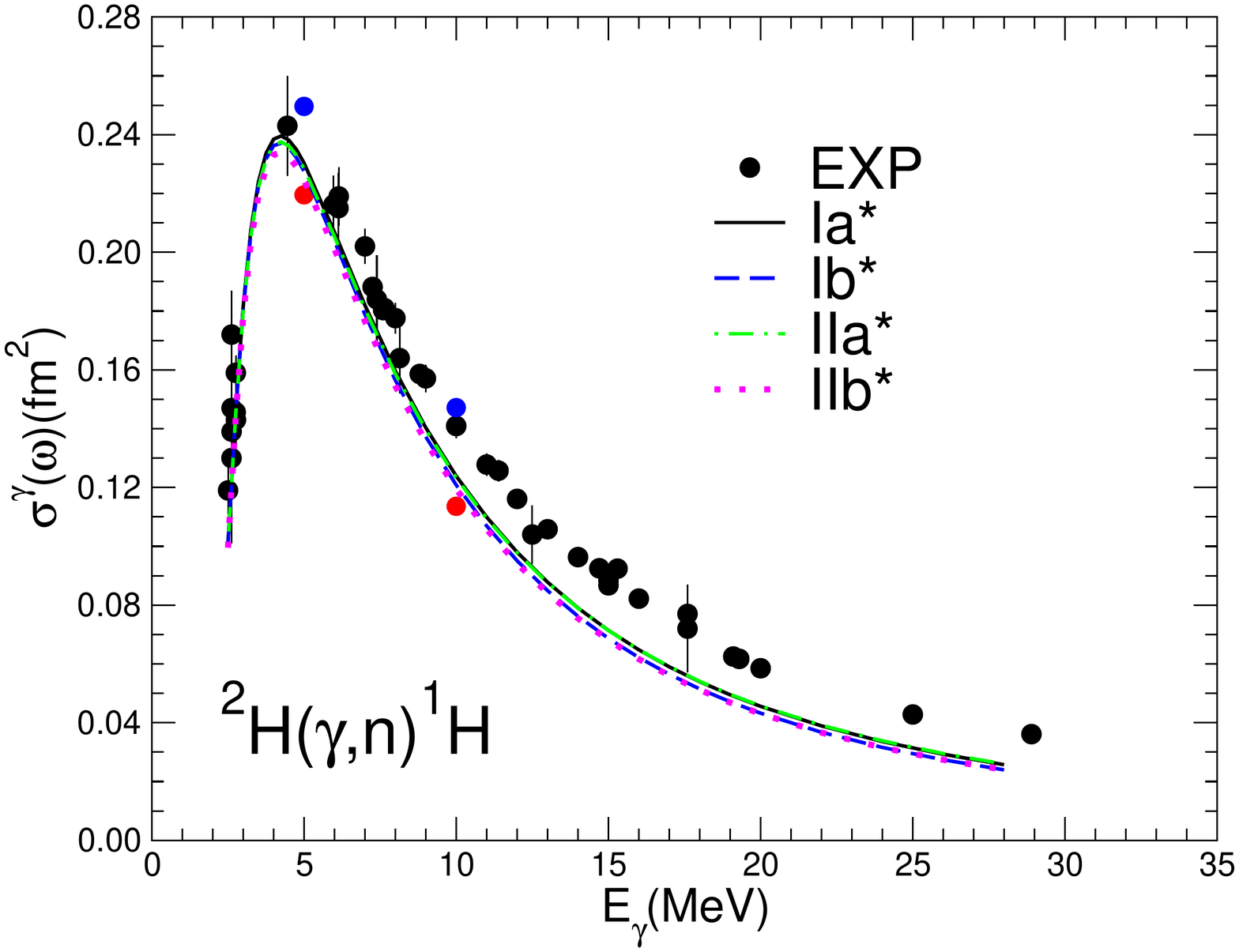}\includegraphics[width=3.75in]{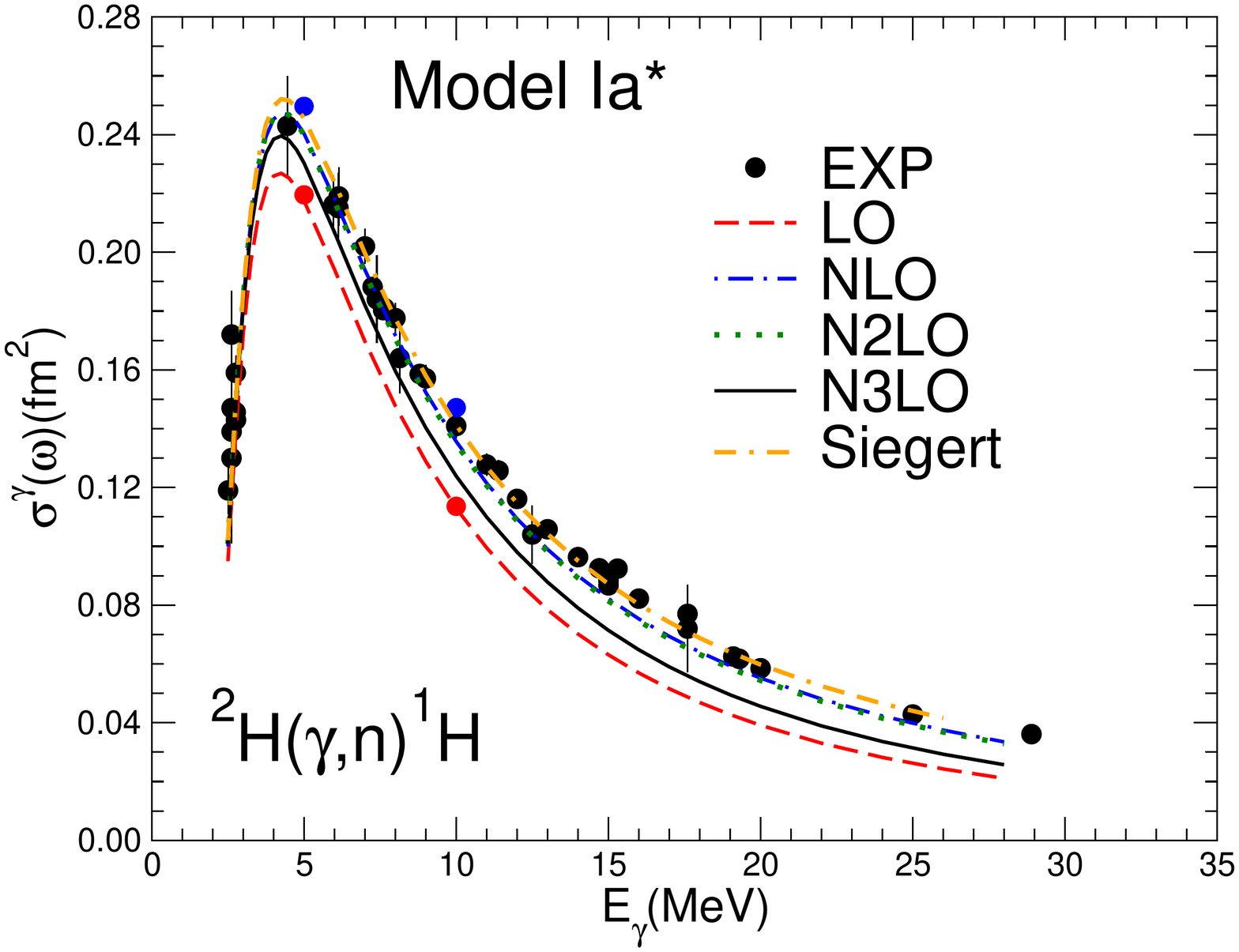}
\caption{Left panel: The low-energy deuteron photodisintegration data (black, blue, and red filled circles)
are compared to predictions
obtained with models Ia$^*$,  Ib$^*$,  IIa$^*$,  and IIb$^*$, including terms up to N3LO in the
current operator.  Right panel: Cumulative contributions to the low-energy deuteron photodisintegration
cross section, obtained at LO, NLO, N2LO, and N3LO for models Ia$^*$, are compared to experimental data;
also shown are the results obtained with the Siegert form of the $E1$ transition operator.}
\label{fig:fdg}
\end{figure}

At low energies, the photodisintegration process is dominated
by the contributions of electric dipole ($E1$) and, to a much
less but still significant extent, electric quadrupole ($E2$)
transitions, connecting the deuteron to the $np$ $^3$P$_J$ states
with $J\,$=$\,0,1,2$ and $^3$S$_1$--$^3$D$_1$ states, respectively (see,
for example, Ref.~\cite{Schiavilla:2005}).
As shown in the left panel of Fig.~\ref{fig:fdg}, the cross sections
obtained with models Ia$^*$/b$^*$ and IIa$^*$/b$^*$ are weakly
dependent on the set of cutoff radii $(R_{\rm S},R_{\rm L})$
regularizing the two-nucleon interactions and currents, and are
systematically lower than the experimental data.

In the right panel of Fig.~\ref{fig:fdg}, the cumulative contributions at LO, NLO, N2LO,
and N3LO are displayed for model Ia$^*$.  The N2LO corrections to this observable
are found to be negligible (the NLO and N2LO curves lay almost on top of each other).
This is because the $\Delta$-excitation current---the leading among
the terms at N2LO---has magnetic dipole character, and therefore primarily
contributes to magnetic dipole ($M1$) transitions, whose strength is much
suppressed relative to $E1$ in the energy regime of interest here.  Of course,
$M1$ transitions become important at threshold and a few-hundred keV's above
threshold; at higher energy, they play a role in polarization observables, as in
$P^\prime_y$---the neutron induced polarization---which in fact results from
interference of $E1$ and $M1$ transitions~\cite{Schiavilla:2005} (unfortunately,
experimental data on this observable in the few MeV range have large
uncertainties, and for this reason $P^\prime_y$ has not been studied
in the present work).
\begin{figure}[bth]
\includegraphics[width=3.75in]{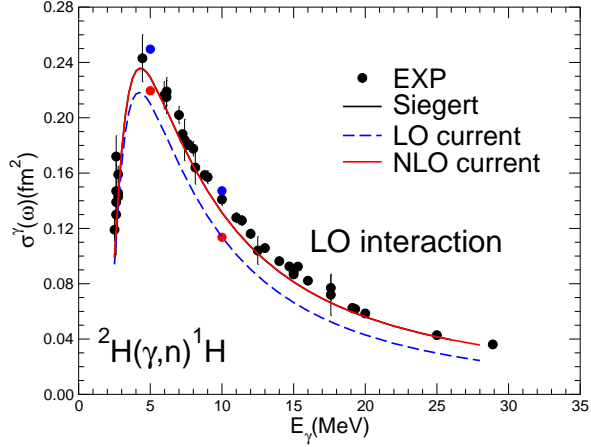}
\caption{The low-energy deuteron photodisintegration
cross section, obtained with the LO interaction of Ref.~\cite{Piarulli:2016} and
currents at LO and up to NLO; also shown are the results obtained with the Siegert form
of the $E1$ transition operator: the curves labeled ``Siegert'' and ``NLO current'' overlap.
The data are represented by the (black, blue, and red) filled circles.}
\label{fig:fdgnlo}
\end{figure}

Of course, $E1$ strength can also be calculated by making use of
the Siegert form for the associated transition operator.   Because of the way the
calculations are carried out in practice (see Ref.~\cite{Schiavilla:2005} for a summary of the methods),
this is most easily implemented by exploiting the identity~\cite{Schiavilla:2004}
\begin{equation}
\label{eq:sieg}
{\bf j}({\bf q})={\bf j}({\bf q})-{\bf j}({\bf q}=0)+i\, \left[ H \,  ,\, \int d{\bf x} \, {\bf x}\, \rho({\bf x}) \right] \ ,
\end{equation}
where $H$ is the nuclear Hamiltonian and $\rho({\bf x})=\delta({\bf x}-{\bf r}_i)\, \epsilon_i+
\delta({\bf x}-{\bf r}_j)\, \epsilon_j$ is the charge density
operator\footnote{There are a number of higher-order corrections to the charge density from
one-body spin-orbit and two-body OPE and TPE as well as center-of-energy terms~\cite{Pastore:2011,Schiavilla:2005}.
These corrections are neglected in the present analysis, since they turn out to
be numerically very small~\cite{Schiavilla:2005}.} ($\epsilon_i$ is the proton projection operator introduced earlier).
Hence, in evaluating matrix elements between the initial deuteron state and final $np$
scattering state, the commutator term simply reduces to
\begin{equation}
\label{eq:sieg1}
i\, \left[ H \,  ,\, \int d{\bf x} \, {\bf x}\, \rho({\bf x}) \right]  \longrightarrow i\, q\, \sum_i \epsilon_i \, {\bf r}_i \ ,
\end{equation} 
where $q$ is the photon energy.  It should be emphasized that the identity above assumes that the current ${\bf j}({\bf q})$
is conserved.  The results of the calculation based on the r.h.s.~of Eq.~(\ref{eq:sieg}) are shown
in the right panel of Fig.~\ref{fig:fdg} (curve labeled Siegert): they are in agreement with data, thus
suggesting that the discrepancies seen at N3LO arise because of the lack of current conservation. 

In order to corroborate this interpretation, we have carried out a calculation
using the interaction at LO (OPE plus LO contact terms proportional to $C_S$ and
$C_T$) fitted to the deuteron binding energy and two-nucleon scattering data up to
energies of 125 MeV, first row of Table I in Ref.~\cite{Piarulli:2016}.
With this interaction, the current including the LO and NLO terms of
Eqs.~(\ref{eq:jlo}) and~(\ref{eq:nlor2}) is exactly conserved in the low $q$ limit.
Indeed at $q\,$=$\,$0, specifically for the NLO current we have
\begin{eqnarray}
 {\bf j}^{\rm NLO}(0)=-\frac{g_A^2}{48\, \pi} \, \frac{m^2_\pi}{f_\pi^2}
 \left({\bm \tau}_i \times {\bm \tau}_j\right)_z \, \hat{\bf r}_{ij}\,\,  C_{R_{\rm L}}(r)\,\,{\rm e}^{-\mu}
 \left[ \left( \frac{3}{\mu^2}+\frac{3}{\mu}+1\right) S_{ij}
 + {\bm \sigma}_i \cdot {\bm \sigma}_j \right] \ ,
\end{eqnarray}
and up to linear terms in the momentum transfer ${\bf q}\cdot  {\bf j}^{\rm NLO}(0) = \left[ v^{\rm LO}_{ij}\, ,\, \rho({\bf q})\right]$,
where $v^{\rm LO}_{ij}$ and $\rho({\bf q})= {\rm e}^{i {\bf q}\cdot {\bf r}_i}\, \epsilon_i 
+  {\rm e}^{i {\bf q}\cdot {\bf r}_j}\, \epsilon_j$
are, respectively, the LO interaction and Fourier transform of the LO charge operator introduced above.
The results of this calculation are displayed in Fig.~\ref{fig:fdgnlo}.  As expected, the curves
labeled ``NLO current'' and ``Siegert'' overlap each other.

Strict adherence to current conservation in the presence of N3LO corrections to the
interaction requires going up to N5LO in the derivation of the electromagnetic operators~\cite{Marcucci:2016}, a rather
daunting task (even in a $\chi$EFT framework with nucleons and pions only, that ignores
$\Delta$ isobars).  It is also unclear how many new LECs would enter, in addition
to the current three, at that order; if there were to be too many, this would obviously reduce
substantially the predictive power of the theory, since there is only a limited number of electromagnetic
observables in the few-nucleon systems (including single nucleons) to constrain these LECs.

\subsection{Deuteron threshold electrodisintegration at backward angles}
The dominant component of the cross section for deuteron electrodisintegration
near threshold at backward angles is the $M1$ transition between the bound deuteron
and the $^1$S$_0$ scattering state~\cite{Hockert:1973}.  As is well known, at large values
of $Q$ this transition rate is dominated by two-body current contributions.  The corrections
from higher partial waves in the final $np$ scattering state are significantly smaller.  Here
we take into account all partial waves in the final state, with full account of the strong
interaction in relatives waves with $J \leq 5$ ($J$ is the total angular momentum)~\cite{Schiavilla:1991}. 
Final-state interaction effects in higher partial waves have been found
to be numerically negligible. 
\begin{figure}[bth]
\includegraphics[width=3.65in]{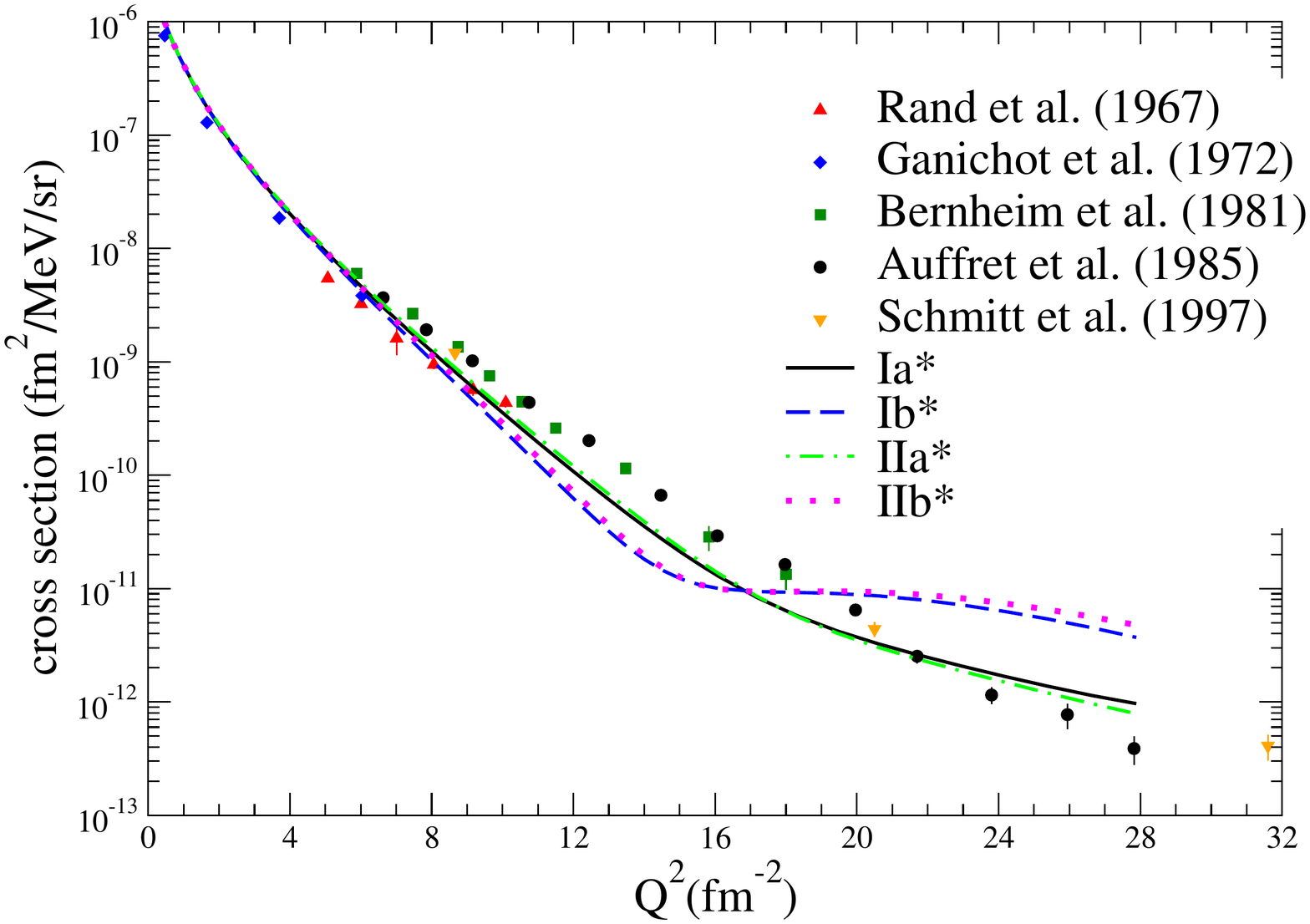}\includegraphics[width=3.65in]{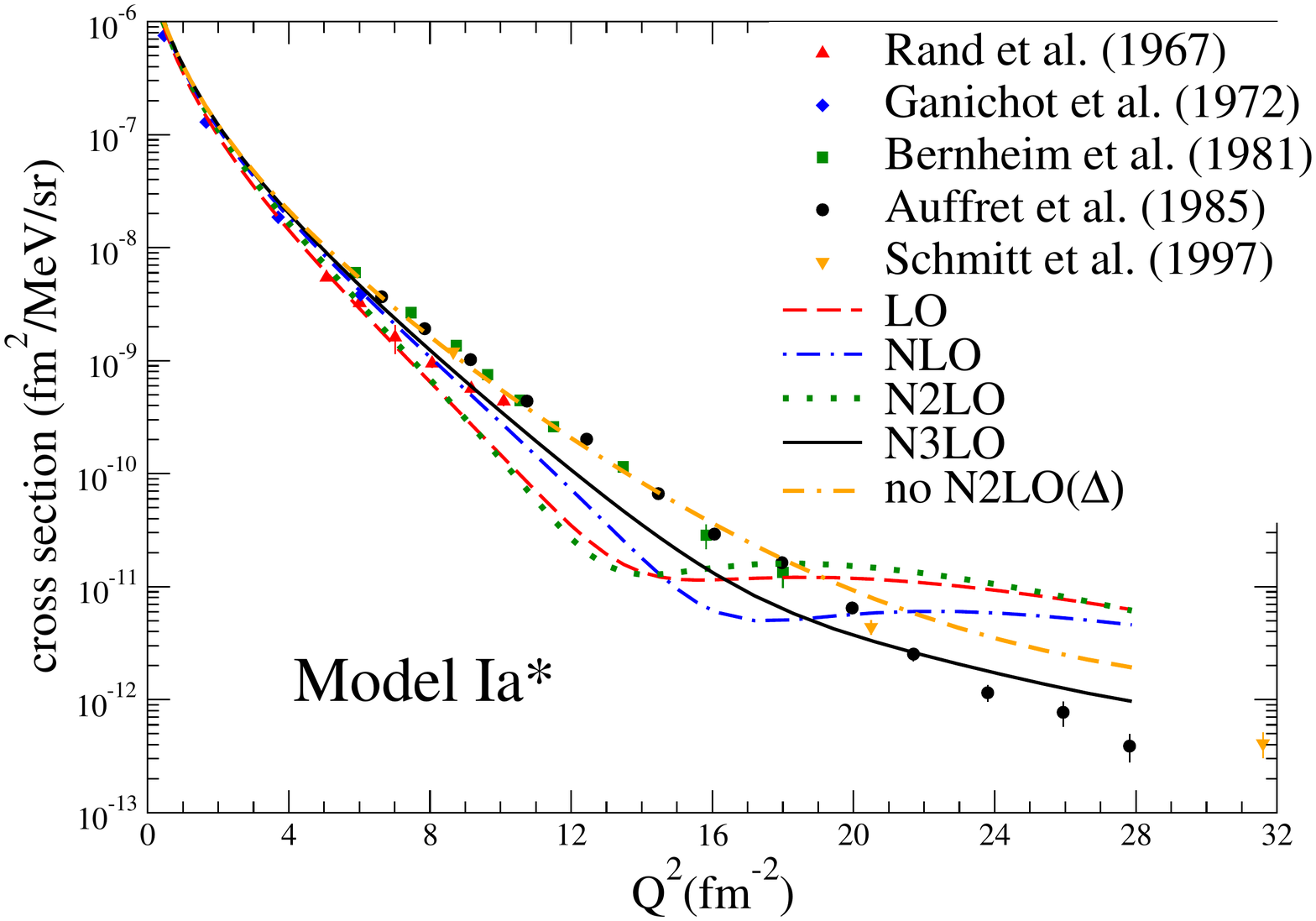}
\caption{Left panel: The deuteron threshold electrodisintegration data at backward angles are
compared to predictions obtained with interaction models Ia$^*$,  Ib$^*$,  IIa$^*$, 
and IIb$^*$, including terms up to N3LO in the current operator.  Right panel: Cumulative
contributions obtained with currents at LO, NLO, N2LO, and N3LO for model Ia$^*$;
also shown are the results at N3LO but excluding the N2LO($\Delta$) current.}
\label{fig:fdpn}
\end{figure}

In Fig.~\ref{fig:fdpn} we compare the calculated cross sections for backward electrodisintegration
with the experimental values.  While the data have been averaged over the interval 0--3 MeV
of the final $np$-pair center-of-mass energy, the theoretical results have been computed
at a fixed energy of 1.5 MeV.  It is known that the effect of the width of the energy
interval over which the cross section values are averaged is very small~\cite{Schiavilla:1991}.
The data, while well reproduced at low values of $Q^2$, are at variance
with theory as $Q^2$ increases beyond 10 fm$^{-2}$, particularly for the set of
harder cutoffs $(R_{\rm S},R_{\rm L})=(0.7,1.0)$ fm of the b models of interactions
and currents.  There is a large cutoff dependence (left panel of Fig.~\ref{fig:fdpn}) 
for $Q\,$$\gtrsim \,$10 fm$^{-2}$, perhaps not surprisingly given that the largest 
four-momentum transfers are comparable, and in fact exceed, the hard scale
$\Lambda_\chi$, below which the $\chi$EFT framework is well defined.  Nevertheless,
it is interesting to note that the trend seen in the present calculations, even at these
high $Q^2$'s, is quantitatively similar to that obtained with interactions and currents
derived from meson-exchange phenomenology, see Refs.~\cite{Carlson:1998,Schiavilla:1991}.

Cumulative contributions at LO, NLO, N2LO, and N3LO are illustrated for model Ia$^*$
in the right panel of Fig.~\ref{fig:fdpn}.  The sharp zero in the LO results seen when only
the $^1$S$_0$ $np$ final state is included~\cite{Schiavilla:1991}, resulting from destructive
interference between the transitions to this state from the S- and D-wave components of the
deuteron, is filled in by the contributions of higher partial waves.  As in the case of the
trinucleon isovector magnetic form factor, for $Q^2\,$$\gtrsim\,$2 fm$^{-2}$ the signs of
the contributions of the currents at LO and NLO, and at N2LO($\Delta$), are opposite, and
the resulting cancellation significantly worsens the agreement between theory and experiment,
see curve labeled ``no N2LO($\Delta$)'' in the right panel of Fig.~\ref{fig:fdpn}.
\section{Conclusions}
\label{sec:concl}
We have carried out a study of magnetic structure and response of the deuteron and
trinucleons with chiral interactions and electromagnetic currents including $\Delta$
intermediate states.  These interactions and currents have been derived and regularized
in configuration space.  While $\Delta$ contributions from leading and sub-leading
$\pi N\Delta$ couplings are accounted for in the TPE component of the two-nucleon
interaction~\cite{Piarulli:2015}, only those at tree level from leading $\gamma N\Delta$
and $\pi N\Delta$ couplings are retained in the OPE and TPE components of, respectively,
the two-body current and three-nucleon interaction~\cite{Piarulli:2018}.  The predicted
magnetic form factors of deuteron and trinucleons are in excellent agreement with
experimental data for momentum transfers corresponding to 3--4 pion masses, and
exhibit a rather weak cutoff dependence in this range, as reflected by the differences
between interaction models a and b.  The present results confirm those of earlier 
studies~\cite{Piarulli:2013}, based on chiral interactions~\cite{Entem:2003,Machleidt:2011}
and currents formulated in momentum space, and which did not include explicitly $\Delta$
degrees of freedom, albeit the cutoff variation here appears to be significantly reduced
when compared to that seen in Ref.~\cite{Piarulli:2013}, particularly at larger values
of momentum transfers.  In this higher momentum-transfer region, the calculations
overestimate the observed deuteron magnetic form factor by an order of magnitude,
and in particular do not reproduce the zero seen in the data at $Q\,$$\simeq$$\,$7 fm$^{-1}$.
In this region the dominant contributions are from the non-minimal
contact current and, in the case of the a models, the isoscalar OPE current.  These
currents are proportional to the LECs $d_1^S$ and $d_2^S$ which have been fixed
by properties at vanishing momentum transfer (the magnetic moments $\mu_d$ and $\mu^S$),
and include hadronic electromagnetic form factors, which we have taken, respectively,
as $G_E^S(Q^2)$---arbitrarily---and as $G_{\gamma\pi\rho}(Q^2)$ by assuming
vector dominance.  So for these reasons, predictions at the higher $Q^2$ should be viewed
as rather uncertain, even after setting aside justified concerns one might
have about the validity of the present $\chi$EFT framework in a regime
where the momentum transfer exceeds the hard scale $\Lambda_\chi\,$$\simeq\,$1 GeV.

The zeros in the magnetic form factors of $^3$He and $^3$H are shifted to lower
momentum transfer than observed.  Thus, the description of the experimental data
at these larger momentum transfers remains problematic, a difficulty already exhibited
by previous $\chi$EFT calculations~\cite{Piarulli:2013} as well as by older calculations
based on meson-exchange phenomenology~\cite{Marcucci:1998}.  This discrepancy
is primarily in the isovector combination of the trinucleon form factors, and is likely to
have its origin in a somewhat too weak overall strength of the isovector component
of the electromagnetic current at large momentum transfer.  By contrast, the calculated
isoscalar combination is very close to data over the whole range of momentum transfers
considered.  It is interesting to note in this connection that the N2LO($\Delta$)
contribution associated with the (isovector) $\Delta$-excitation current is comparable in
magnitude to, and of opposite sign than, the NLO contribution in the diffraction region, see
lower right panel of Fig.~\ref{fig:fmcbt}.  This destructive interference then appears to be
the culprit of the current failure of theory to reproduce experiment.  Similar considerations
also apply to the deuteron threshold electrodisintegration cross sections at backward angles,
which are underestimated by theory for $Q^2\,$$\gtrsim\,$8 fm$^{-2}$---a failure known to
occur in the older meson-exchange calculations too (see Ref.~\cite{Carlson:1998} and
references therein).

The sign change of the contribution associated with the $\Delta$-excitation current
also occurred in the Marcucci {\it et al.}~calculations of trinucleon form factors~\cite{Marcucci:1998},
albeit at a larger momentum transfer than here ($Q\,$$\simeq\,$3 fm$^{-1}$ versus $\simeq\,$
1.5 fm$^{-1}$), presumably due to the much harder cutoff adopted in that work. In the
Marcucci {\it et al.}~paper, contributions which in a $\chi$EFT approach would approximately
correspond to TPE terms with $\Delta$ intermediate states  in two- and three-body currents,
were also considered, and found to have the same sign as, but to be suppressed by an order
of magnitude relative to, those from the OPE current.  Whether this estimate of $\Delta$ effects
at the TPE level remains valid in the present $\chi$EFT framework is an open question we
hope to address in the future. 

Finally, we have shown that the inability of theory to provide a satisfactory description of
the measured deuteron photodisintegration cross sections at low energies originates from
the lack of current conservation.  This flaw can be corrected, at least in processes in which
electric-dipole transitions are dominant, by making use of the Siegert form of the $E1$
operator.  In practical calculations this is most easily implemented via the replacements
in the r.h.s.~of Eqs.~(\ref{eq:sieg})--(\ref{eq:sieg1}).

We conclude by noting that the present work complements that of Ref.~\cite{Baroni:2018}.
Together, these two papers provide the complete set of chiral electroweak currents at one
loop with fully constrained LECs for use with the local chiral interactions developed in
Ref.~\cite{Piarulli:2016} (models Ia/b and IIa/b) and Ref.~\cite{Baroni:2018} (models
Ia$^*$/b$^*$ and IIa$^*$/b$^*$), thus opening up the possibility to study radiative and
weak transitions in systems with mass number $A>4$ with QMC methods, and
low-energy neutron and proton radiative captures on deuteron and $^3$H/$^3$He, and
proton weak capture on $^3$He (the Hep process) with HH techniques. Work along these
lines is in progress.
 
\vspace{0.5cm}
One of the authors (R.S.) thanks the T-2 group in the Theoretical Division at LANL,
and especially J.\ Carlson and S.\ Gandolfi, for the support and warm hospitality extended to him
during a sabbatical visit in the Fall 2018, when part of this research was completed.  The support
of the U.S.~Department of Energy, Office of Science,  Office of Nuclear Physics, under contracts
DE-AC05-06OR23177 (R.S.) and DE-AC02-06CH11357 (A.L.,~M.P.,~S.C.P., and~R.B.W.), and
award DE-SC0010300 (A.B.), is gratefully acknowledged. The work {of~A.L.,~S.P.,~M.P.,~S.C.P.,
and~R.B.W.} has been further supported by the NUclear Computational Low-Energy Initiative
(NUCLEI) SciDAC project.  Computational resources provided by the National Energy Research
Scientific Computing Center (NERSC) are also thankfully acknowledged.

\appendix
\section{Loop corrections to the electromagnetic current in configuration space}
\label{app:a1}
In this appendix we sketch the derivation of the configuration-space expressions for
the loop corrections to the electromagnetic current.  At low momentum transfer (denoted as {\bf q}),
these corrections read in momentum space~\cite{Pastore:2009,Piarulli:2013}
\begin{equation}
\label{eq:jloopk}
{\widetilde{\bf j}}^{\,{\rm N3LO}}_{\, \rm TPE}({\bf k}_{ij})= i\,\tau_{j,z}
\left[\widetilde{F}_0(k_{ij}) \,{\bm \sigma}_i - \widetilde{F}_2(k_{ij})\, \frac{{\bf k}_{ij}\,{\bm \sigma}_i\cdot{\bf k}_{ij}}{k_{ij}^2}\right] \times \frac{ {\bf q}}{2\, m_\pi}  
-i\, ({\bm \tau}_i\times {\bm \tau}_j)_z \,{\bm \nabla}_{{\! k}_{ij}}\,\widetilde{F}_1(k_{ij}) +
(i \rightleftharpoons j) \ ,
\end{equation}
where we have defined
\begin{equation}
{\bf k}_{ij}=({\bf k}_i-{\bf k}_j)/2 \ , \qquad {\bf q}={\bf k}_i+{\bf k}_j \ ,
\end{equation}
and
\begin{eqnarray}
\widetilde{F}_0(k)&=&\frac{g_A^2}{128\,\pi^2}\,\frac{2\, m_\pi}{f_{\pi}^4}\,\left\{
1 -2\, g_A^2+\frac{  8\,g_A^2\, m_\pi^2 }{k^2+4\, m_\pi^2}
+G(k)\left[ 2-2\, g_A^2
-\frac{  4\,(1+g_A^2)\, m_\pi^2 }{k^2+4\, m_\pi^2} 
+\frac{16\, g_A^2 \,m_\pi^4 }{(k^2+4\, m_\pi^2)^2} \right] \right\}\ ,
\label{eq:f0k} \\
\widetilde{F}_1(k)&=&\frac{1}{1536 \, \pi^2\,f_{\pi}^4}\,
G(k) \bigg[4 m_{\pi}^2(1+4 g_A^2-5 g_A^4)+k^2(1+10 g_A^2 - 23 g_A^4)
-\frac{48\, g_A^4 m^4_\pi}
{4\,  m^2_\pi+k^2}\bigg] \ ,
\label{eq:f1k}\\
\widetilde{F}_2(k)&=&\frac{g_A^2}{128\,\pi^2}\, \frac{2\, m_\pi}{f_{\pi}^4}\,\left\{
2-6\, g_A^2+ \frac{  8\,g_A^2\, m_\pi^2 }{k^2+4\, m_\pi^2}
+G(k) \left[4\, g_A^2
-\frac{  4\,(1+3\, g_A^2)\, m_\pi^2 }{k^2+4\, m_\pi^2} 
+\frac{16\, g_A^2 \,m_\pi^4 }{(k^2+4\, m_\pi^2)^2} \right] \right\}\ ,
\label{eq:f2k}
\end{eqnarray}
with the loop function $G(k)$ given by
\begin{equation}
G(k)=\frac{\sqrt{4\,m_{\pi}^2+k^2}}{k}\ln 
\frac{\sqrt{4\,m_{\pi}^2+k^2}+k}{\sqrt{4\,m_{\pi}^2+k^2}-k} \ .
\label{eq:loopf}
\end{equation}
We note that the term $\widetilde{F}_1(k)$ is related to the
TPE interaction ${\bm \tau}_i\cdot{\bm \tau}_j \,\widetilde{ v}^{\,{\rm NLO}}_{\,2\pi}(k)$
(in a $\chi$EFT with nucleon and pion degrees of freedom only) via
$\widetilde{F}_1(k)=   \widetilde{v}^{\,{\rm NLO}}_{\,2\pi}(k)/2$,
and that the longitudinal term proportional to $\widetilde{F}_1(k)$ satisfies
current conservation with this interaction (in the limit of small $q$), namely
\begin{equation}
\left[ \, \widetilde{v}^{\,{\rm NLO}}_{\, 2\pi}(|{\bf k}_{ij}-{\bf q}/2|)\,
 {\bm \tau}_i\cdot{\bm \tau}_j \,\, , \,\, \frac{1+\tau_{i,z}}{2}\, \right] 
+ (i\rightleftharpoons j) = -i\, ({\bm \tau}_i \times{\bm \tau}_j)_z\,
 {\bf q}\cdot{\bm \nabla}_{{\bf k}_{ij}} \, \widetilde{F}_1(k_{ij})+ (i\rightleftharpoons j) \ .
\end{equation} 
 In configuration space, we obtain
\begin{eqnarray}
\label{eq:jloopr}
{\bf j}^{\,{\rm N3LO}}_{\,\rm TPE}({\bf q})&=&
i\,\tau_{j,z}\,\, {\rm e}^{i{\bf q}\cdot {\bf R}_{ij}}
\left[ {\bm \sigma}_i \int_{\bf k} {\rm e}^{i{\bf k}\cdot {\bf r}_{ij}} \widetilde{F}_0(k)
+ \left(2\,m_\pi\right)^2 \left({\bm \sigma}_i\cdot{\bm \nabla}^{{\lambda}_{ij}}\right)\, {\bm \nabla}^{{\lambda}_{ij}}
 \int_{\bf k} {\rm e}^{i{\bf k}\cdot {\bf r}_{ij}}\,\, \frac{\widetilde{F}_2(k)}{k^2}\,  \right] 
 \times \frac{{\bf q}}{2\, m_\pi} \nonumber\\
&&-\frac{1}{2} ({\bm \tau}_i\times {\bm \tau}_j)_z\,\,  {\rm e}^{i{\bf q}\cdot {\bf R}_{ij}} \,{\bf r}_{ij}\,
v_{\,2\pi}^{\rm NLO}(r_{ij})+ (i \rightleftharpoons j) \ , 
\end{eqnarray}
where ${\bm \nabla}^{{\lambda}_{ij}}={\partial}/{\partial {\bm \lambda}_{ij}}$ (${\bm \lambda}_{ij}\,$=$\, 2\, m_\pi\,{\bf r}_{ij}$)
and $\int_{\bf k} =\int \! {d{\bf k}}/{(2\pi)^3}$.
The Fourier transforms of $\widetilde{F}_0(k)$ and $\widetilde{F}_2(k)/k^2$
in terms of $x=k/\left(2\, m_\pi\right)$ reduce to (dropping the subscripts $ij$)
\begin{eqnarray}
\int_{\bf k} {\rm e}^{i{\bf k}\cdot{\bf r}} \, \widetilde{F}_0(k) &=& \frac{(2\, m_\pi)^3}{2\, \pi^2} \,\frac{1}{\lambda} \int_0^\infty dx \, x\, {\rm sin} (x\lambda)\, \widetilde{F}_0(x) \ , \\
\left(2\, m_\pi\right)^2 \int_{\bf k} {\rm e}^{i{\bf k}\cdot{\bf r}} \, \frac{\widetilde{F}_2(k)}{k^2} &=& \frac{\left(2\, m_\pi\right)^3}{2\, \pi^2} \,\frac{1}{\lambda} \int_0^\infty dx \, \frac{{\rm sin} (x\lambda)}{x}\, \widetilde{F}_2(x) \ . 
\end{eqnarray}
In order to carry out the sine transforms above, we find it convenient to express
$G(x)$ in Eq.~(\ref{eq:loopf}) as
\begin{equation}
G(x)=2+G^\star(x)\ ,\qquad G^\star(x) = \int_0^1 dz\, {\rm ln} \left[1+x^2\left(1-z^2\right)\right] \ ,
\end{equation}
and $G^\star(x)$ diverges logarithmically in the limit $|x|\gg 1$.
In terms of the variable $x$, the functions $F_i(x)$ are written as
\begin{equation}
\frac{\left(2\,m_\pi\right)^3} {2\, \pi^2}\widetilde{F}_i(x)=
A_i(x)+G^\star(x) \,B_i(x) \ ,
\end{equation}
where
\begin{eqnarray}
 A_0(x)&=&\frac{g_A^2}{256\,\pi^4}\,\frac{\left(2\, m_\pi\right)^4}{f_{\pi}^4}
\left[ 5-6\, g_A^2 -\frac{2}{x^2+1} +\frac{2\, g_A^2}{\left(x^2+1\right)^2} \right] \ ,\\
 B_0(x)&=&\frac{g_A^2}{256\,\pi^4}\,\frac{\left(2\, m_\pi\right)^4}{f_{\pi}^4}
\left[2-2\, g_A^2 -\frac{1+g_A^2}{x^2+1} +\frac{g_A^2}{\left(x^2+1\right)^2} \right]  \ ,\\
A_2(x)&=&\frac{g_A^2}{128\,\pi^4}\,\frac{\left(2\, m_\pi\right)^4}{f_{\pi}^4}
\left[ 1+g_A^2 -\frac{1+2\,g_A^2}{x^2+1} +\frac{g_A^2}{\left(x^2+1\right)^2} \right] \ ,\\
 B_2(x)&=&\frac{g_A^2}{256\,\pi^4}\,\frac{\left(2\, m_\pi\right)^4}{f_{\pi}^4}
\left[4\, g_A^2 -\frac{1+3\,g_A^2}{x^2+1} +\frac{g_A^2}{\left(x^2+1\right)^2} \right]  \ .
\end{eqnarray}
We then obtain
\begin{eqnarray}
\int_{\bf k} {\rm e}^{i{\bf k}\cdot{\bf r}} \, \widetilde{F}_0(k) 
 &=&\frac{1}{2\,\lambda} \int_{-\infty}^\infty dx \, x\, {\rm sin} (x\lambda)
 \left[A_0(x)+G^\star(x)\, B_0(x)\right]\ , \\
\left(2\, m_\pi\right)^2 \int_{\bf k} {\rm e}^{i{\bf k}\cdot {\bf r}}\,\, \frac{\widetilde{F}_2(k)}{k^2} &=& 
 \,\frac{1}{2\,\lambda}\int_{-\infty}^\infty dx\, \frac{ {\rm sin}(x\lambda)}{x}\left[A_2(x)+G^\star(x)\, B_2(x)\right]\ ,
 \end{eqnarray}
where the integration limits over $x$ have been extended to the range $(-\infty,\infty)$.
The function $F_0(x)$ diverges in the limit $x \longrightarrow \infty$ and must be regularized
before the sine transform can be carried out.  To this end, we define
\begin{eqnarray}
A_0^\infty(x) &=& \frac{g_A^2}{256\,\pi^4}\,\frac{\left(2\, m_\pi\right)^4}{f_{\pi}^4}\left( 5-6\, g_A^2 \right) \ ,\\
B^\infty_0(x)&=&\frac{g_A^2}{256\,\pi^4}\,
 \frac{\left(2\, m_\pi\right)^4}{f_{\pi}^4}\left(2-2\,g_A^2\right) \ ,
 \end{eqnarray} 
and then choose to subtract from $F_0(x)$ its asymptotic behavior proportional to $A^\infty_0(x)+G^\star(x) B_0^\infty(x)$,
namely
 \begin{eqnarray}
\frac{\left(2\,m_\pi\right)^3}{2\, \pi^2}  \overline{F}_0(x)&=& 
\frac{\left(2\,m_\pi\right)^3}{2\, \pi^2}\left[  \widetilde{F}_0(x)-
 F_0^\infty(x)\right] 
=\overline{A}_0(x)+G^\star(x) \,\overline{B}_0(x)\ ,
 \end{eqnarray}
where
 \begin{eqnarray}
\overline{A}_0(x)  &=&-\frac{g_A^2}{128\,\pi^4}\,\frac{\left(2\, m_\pi\right)^4}{f_{\pi}^4}
\frac{1}{x^2+1} \left(1 -\frac{g_A^2}{x^2+1} \right) \ ,\\
\overline{B}_0(x)&=& -\frac{g_A^2}{256\,\pi^4}\,\frac{\left(2\, m_\pi\right)^4}{f_{\pi}^4}
 \frac{1}{x^2+1} \left(1+g_A^2 -\frac{g_A^2}{x^2+1} \right) \ .
\end{eqnarray}
The Fourier transform of $F_0^\infty(k)$ is then regularized by multiplication of a Gaussian cutoff
${\rm exp}\left(-k^2 R^2_{\rm S}/4\right)$ as in Eq.~(\ref{eq:e221}) to obtain ($z\,$=$\, r/R_{\rm S}$)
\begin{eqnarray}
\label{eq:e28}
\hspace{-1cm}
F^{(0)}(z;\infty) 
&=&\frac{g_A^4}{1024\,\pi^2}\,\frac{\left(2\, m_\pi\right)^4}{f_{\pi}^4} 
\left(\frac{1}{g_A^2}-2\right) C^{(0)}_{R_{\rm S}}(z)  \nonumber\\
&&+\frac{g_A^4}{1024\,\pi^4}\,\frac{\left(2\, m_\pi\right)^4}{f_{\pi}^4} \frac{1}{\left(m_\pi R_{\rm S}\right)^3}
\left(\frac{1}{g_A^2} -1\right)
 \int_0^\infty \!\ dx\, x^2\, j_0(x z)\, G(x/R_{\rm S})\, 
{\rm e}^{-x^2/4} \ ,
\end{eqnarray}
where we have reinstated the function $G(k)$ in the second line of the above equation.
\subsection{Sine transforms}
We collect here the formulae needed for the sine transforms involving $G^\star(x)$ (those without
$G^\star(x)$ are elementary)
\begin{eqnarray}
\label{eq:e35}
&&\hspace{-4.5cm}\int_{-\infty}^\infty dx \, x\, {\rm sin} (x\lambda)\, \frac{G^\star(x)}{x^2+1} 
=\pi \!\! \int_0^1 dz\, \Big[ {\rm e}^{-\lambda} \, 
E_1\left (\lambda\, \alpha_z-\lambda\right)+{\rm e}^{\lambda} \, 
E_1\left(\lambda\,\alpha_z+\lambda\right)\Big] -2\, \pi\, {\rm e}^{-\lambda}\ , \\
\label{eq:e36}
&&\hspace{-4.5cm} \int_{-\infty}^\infty dx \, x\, {\rm sin} (x\lambda)\, \frac{G^\star(x)}{(x^2+1)^2} =
- \pi\,\lambda\,{\rm e}^{-\lambda}+\pi\, {\rm e}^{-\lambda}
 \int_0^1 dz  \,\frac{1}{\alpha_z^2-1}\left[ 1-{\rm e}^{-\lambda(\alpha_z-1)} \right] \nonumber\\
 &&\hspace{0.5cm} +\frac{\pi}{2} \int_0^1dz  \Big[\lambda\,{\rm e}^{-\lambda} \, 
 E_1\left (\lambda\,\alpha_z-\lambda\right)
 -\lambda\,{\rm e}^{\lambda} \, E_1\left (\lambda\, \alpha_z+\lambda\right) \Big] \ ,\\
 \label{eq:e37}
&&\hspace{-4.5cm} \int_{-\infty}^\infty dx\, \frac{ {\rm sin}(x\lambda)}{x}\, G^\star(x)= 2\, \pi\int_0^1dz\, E_1(\lambda\,\alpha_z)  \ ,\\
\label{eq:e38}
&&\hspace{-4.5cm} \int_{-\infty}^\infty dx\, \frac{ {\rm sin}(x\lambda)}{x}\, \frac{G^\star(x)}{x^2+1}=  2\, \pi\left[ 
{\rm e}^{-\lambda}+\int_0^1dz\, E_1\left(\lambda\, \alpha_z\right) \right]
-\,\pi \int_0^1 dz\, \Big[ {\rm e}^{-\lambda} \, 
E_1\left ( \lambda\, \alpha_z-\lambda\right)+{\rm e}^{\lambda} \, 
E_1\left ( \lambda\, \alpha_z+\lambda\right) \Big] \ ,\\
\label{eq:e39}
\hspace{-0.5cm} \int_{-\infty}^\infty dx\, \frac{ {\rm sin}(x\lambda)}{x}\, \frac{G^\star(x)}{(x^2+1)^2}&=& 2 \,\pi
\left( 1+\frac{\lambda}{2}\right){\rm e}^{-\lambda}-
 \pi\,{\rm e}^{-\lambda} \int_0^1 dz  \,\frac{1}{\alpha_z^2-1}\left[ 1-{\rm e}^{-\lambda(\alpha_z-1)} \right] \nonumber\\
&&\hspace{-0.5cm} -\pi \int_0^1dz \left[
\left(1+\frac{\lambda}{2}\right){\rm e}^{-\lambda} \,  E_1\left (\lambda\, \alpha_z-\lambda\right)
+\left(1-\frac{\lambda}{2}\right) {\rm e}^{\lambda} \, E_1\left (\lambda\, \alpha_z+\lambda\right)
-2 \, E_1\left( \lambda\, \alpha_z\right)\right]  \ ,
\end{eqnarray}
where 
\begin{equation}
\alpha_z=\frac{1}{\sqrt{1-z^2}} \ge 1  \ ,
\end{equation}
and we have introduced the exponential integral defined as~\cite{Abramowitz:1972}
\begin{equation}
E_1(x) =\int_x^\infty dt\, \frac{{\rm e}^{-t}}{t}\ ,
\end{equation}
with the following series expansion and asymptotic behavior
\begin{equation}
E_1(x) =-\gamma-{\rm ln}\, x -\sum_{n=1}^\infty (-1)^n\, \frac{x^n}{n\, n!} \ ,\qquad
E_1(x) =\frac{{\rm e}^{-x}}{x}\left( 1-\frac{1!}{x}+\frac{2 !}{x^2}-\frac{3!}{x^3}+\cdots\right) \,\,\,{\rm for} \,\,\, x\gg 1\ ,\end{equation}
and $\gamma$ is Euler's number.

We will now illustrate the evaluation of one of the integrals above, which we
carry out by contour integration in the
complex plane by proceeding in a similar way as in Ref.~\cite{Baroni:2018}. We consider
\begin{eqnarray}
\!\!\!\!\!\!\! \int_{-\infty}^\infty dx \, x\, {\rm sin} (x\lambda)\,
\frac{G^\star(x)}{x^2+a^2} &=&\! \int_0^1 dz \,\, {\rm Im}\int_{-\infty}^\infty dx \, \frac{x}{x^2+a^2}\, {\rm e}^{i\,x\lambda}\,
{\rm ln} \left[1+x^2\left(1-z^2\right)\right] \nonumber\\
\!\!\!\!\!\!\!&=&\!\left[\int_0^1 dz\,   {\rm ln}\left(1-z^2\right) \right]
{\rm Im}\int_{-\infty}^\infty dx \, \frac{x}{x^2+a^2}\, {\rm e}^{i\,x\lambda}+ \int_0^1 dz\,   I(z,\lambda),
\end{eqnarray}
where $a$ is a parameter ($a=1$ is the value of interest) and
\begin{equation}
I(z,\lambda)=  {\rm Im}\int_{-\infty}^\infty dx \, \frac{x}{x^2+a^2}\, {\rm e}^{i\,x\lambda}\,
{\rm ln} \left(\frac{1}{1-z^2}+x^2\right) \ ,
\end{equation}
i.e., we perform the sine transform first and then the parametric integration
over $z$.
\begin{figure}[bth]
 \includegraphics[width=3in]{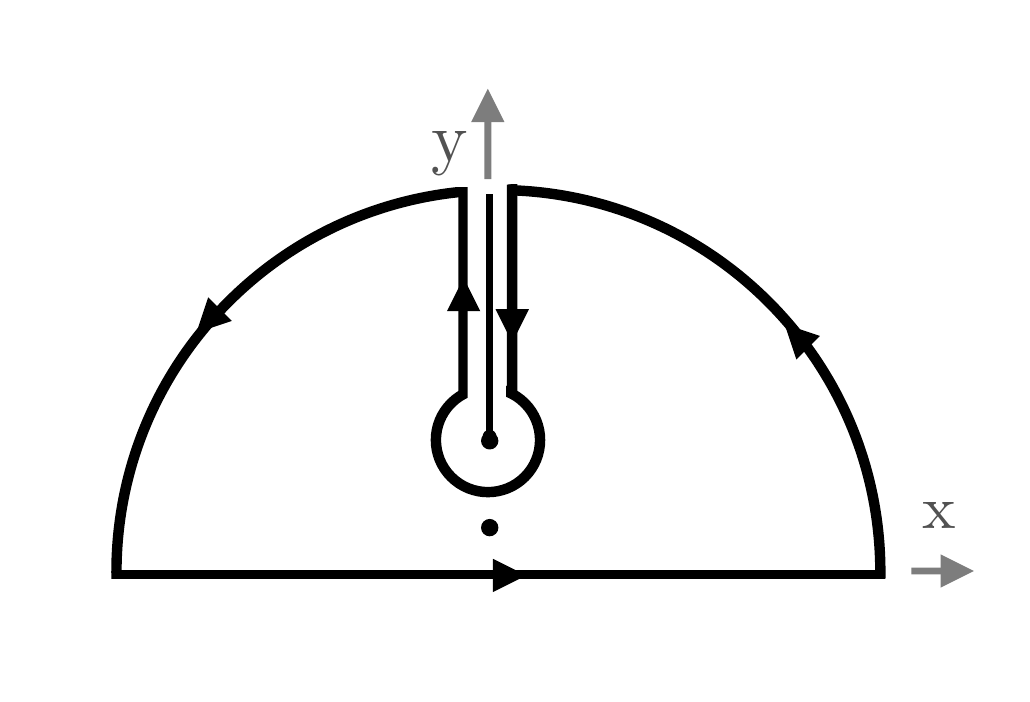}
 \caption{Integration contour.}
\label{fig:f1}
\end{figure}
We define the function of the complex variable $\eta$
\begin{equation}
f(\eta)={\rm e}^{i\,\eta \lambda} \,\frac{\eta}{\eta^2+a^2}
\,\,{\rm ln} \left[ (\eta-i\,\alpha_z)(\eta+i\,\alpha_z)\right] \ .
\end{equation}
This function has branch points at $\eta=\pm \,i\, \alpha_z $ and simple poles
at $\eta=\pm\, i\, a$ ($0< a \leq 1$), but is otherwise analytic.  The upper cut is taken
from $i\, \alpha_z$ to $+i\,\infty$ (along the positive imaginary axis), while the lower one from
$-i\, \alpha_z$ to $-i\, \infty$ (along the negative imaginary axis). We consider the closed contour $C$ as in Fig.~\ref{fig:f1},
so that
\begin{equation}
\oint_C d\eta\, f(\eta) =2\pi\,i\, {\rm Res}\, f(\eta)\mid_{\eta=i\, a}  \ .
\label{eq:e10}
\end{equation}
Before evaluating the integral above, we need to consider the values of $f(\eta)$ to the right and left
of the cut running along the positive imaginary axis.  To this end, we define
\begin{equation}
\eta-i\,\alpha_z=r_+\, {\rm e}^{i\,\theta_+} \,\,\, {\rm with}\,\,\, -\frac{3\,\pi}{2}\leq  \theta_+ \leq \frac{\pi}{2}  \ ,
\qquad \eta+i\,\alpha_z=r_-\, {\rm e}^{i\, \theta_-} \,\,\, {\rm with}\,\,\, -\frac{\pi}{2} \leq  \theta_- \leq \frac{3\,\pi}{2} \ ,
\end{equation}
the restrictions on $\theta_\pm$ ensuring that the cuts are not crossed.  For a given $\eta$,
the difference along the upper cut (corresponding to $\eta=i\, y$ with $y \ge \alpha_z$) is given by
\begin{equation}
{\rm ln}\left(\eta^2+\alpha^2_z\right)\mid_{{\rm left\,\,\,of\,\,\, cut}}
-{\rm ln}\left(\eta^2+\alpha^2_z\right)\mid_{{\rm right\,\,\,of\,\,\, cut}} =-\,2\, \pi\, i \ .
\end{equation}
The contributions of the big arcs of radius $R$
and small circle of radius $r$ around the brach point $+\, i\, \alpha_z$ vanish
as, respectively, $R \rightarrow \infty$ and $r\rightarrow 0$,
while on the segments left and right of the upper cut we find
\begin{equation}
\int_{\rm left} d\eta\, f(\eta)- \int_{\rm right} d\eta\, f(\eta)= -2\, \pi \, i  \int_{\alpha_z}^\infty dy\,  {\rm e}^{-y \,\lambda}\, 
\frac{y}{y^2-a^2}  \ .
\end{equation}
Therefore from Eq.~(\ref{eq:e10}), after evaluating the residue, we arrive at
\begin{equation}
I(z,\lambda)-2\, \pi  \int_{\alpha_z}^\infty dy\,  {\rm e}^{-y \,\lambda}\, \frac{y}{y^2-a^2} 
=\pi \, {\rm e}^{-a\lambda}\,{\rm ln} (\alpha_z^2-a^2) \ ,
\end{equation}
from which we deduce
\begin{eqnarray}
\hspace{-0.5cm} \int_{-\infty}^\infty dx \, x\, {\rm sin} (x\lambda)\, \frac{G^\star(x)}{x^2+a^2} =
2\, \pi  \int_0^1\!\! dz\, \int_{\alpha_z}^\infty \!dy\,  {\rm e}^{-y \,\lambda}\, \frac{y}{y^2-a^2} 
+\,\pi\, {\rm e}^{-a\lambda}  \int_0^1 \!\!dz \left[ {\rm ln}\left(1-z^2\right)+
{\rm ln}\left(\frac{1}{1-z^2}-a^2\right) \right] \ .
\end{eqnarray}
The first term on the r.h.s.~of the equation above can be expressed in terms of exponential
functions by noting that
\begin{eqnarray}
\int_{\alpha_z}^\infty dy\,  {\rm e}^{-y \,\lambda}\, \frac{y}{y^2-a^2}=
 \frac{{\rm e}^{-a\lambda}}{2} \, E_1\left (\lambda\,\alpha_z-\lambda\,a\right)
+\frac{{\rm e}^{a\lambda}}{2} \, 
E_1\left (\lambda\,\alpha_z+\lambda\, a\right)\ ,
\end{eqnarray}
so that
\begin{eqnarray}
\int_{-\infty}^\infty dx \, x\, {\rm sin} (x\lambda)\, \frac{G^\star(x)}{x^2+a^2} 
&=&\pi  \int_0^1 dz\, \Big[ {\rm e}^{-a\lambda} \, 
E_1\left (\lambda\, \alpha_z-\lambda\, a\right)+{\rm e}^{a\lambda} \, 
E_1\left (\lambda\, \alpha_z+\lambda\, a\right) \Big] \nonumber\\
&&+\,2\, \pi\, {\rm e}^{-a\lambda}\left(\frac{\sqrt{1-a^2}}{a} \, {\rm arctan}\frac{a}{\sqrt{1-a^2}} -1\right)\ .
\end{eqnarray}

\subsection{Correlation functions}
Inserting the expressions above, we find the Fourier transforms of $\widetilde{F}_0(k)$ and
$\widetilde{F}_2(k)/k^2$ to be given by
\begin{eqnarray}
\label{eq:e55}
&&F_0^{(0)}(\lambda)\!=\! \int_{\bf k} {\rm e}^{i{\bf k}\cdot{\bf r}} \, \widetilde{F}_0(k) =\frac{g_A^4}{256\,\pi^3}\,
\frac{\left(2\, m_\pi\right)^4}{f_{\pi}^4}\bigg\{
  \frac{{\rm e}^{-\lambda}}{\lambda} \left[ 1+\frac{1}{2}\int_0^1dz\, \frac{1-{\rm e}^{-\lambda(\alpha_z-1)}}{\alpha_z^2-1} \right]
-\left[\left(\frac{1}{g_A^2} +1\right)\frac{1}{\lambda} -\frac{1}{2} \right]  E^{(-)}_1(\lambda) \nonumber\\
&& \hspace{7cm} -\left[ \left( \frac{1}{g_A^2}+1\right)\frac{1}{\lambda}+\frac{1}{2}\right]
 E^{(+)}_1(\lambda)  \bigg\}\ ,\\
&&F^{(0)}_2(\lambda)\!=\! (2\, m_\pi)^2\!\int_{\bf k} {\rm e}^{i{\bf k}\cdot{\bf r}} \, \frac{\widetilde{F}_2(k)}{k^2} = -\frac{g_A^4}{256\,\pi^3}\,\frac{\left(2\, m_\pi\right)^4}{f_{\pi}^4}\bigg\{
  \frac{{\rm e}^{-\lambda}}{\lambda} 
  \left[ 1+\frac{1}{2}\int_0^1dz\, \frac{1-{\rm e}^{-\lambda(\alpha_z-1)}}{\alpha_z^2-1} \right]
-\left[\left(\frac{1}{g_A^2} +2\right)\frac{1}{\lambda} -\frac{1}{2} \right]\!
E^{(-)}_1(\lambda) \nonumber\\
&&\hspace{7cm}- \left[ \left( \frac{1}{g_A^2}+2\right)\frac{1}{\lambda}+\frac{1}{2}\right] \! E^{(+)}_1 (\lambda)  
+ \left(\frac{1}{g_A^2} -2\right) \frac{1}{\lambda} \,
 E_1^{(0)} (\lambda) \bigg\}\ ,
\end{eqnarray}
where we have defined
\begin{equation}
E_1^{(\pm)}(\lambda)=\frac{{\rm e}^{\pm \lambda}}{2} \int_0^1 dz\, E_1\left (\lambda\,\alpha_z\pm \lambda\right) \ ,
\qquad E_1^{(0)}(\lambda)=\int_0^1 dz\, E_1\left (\lambda\,\alpha_z\right) \ .
\end{equation}
The correlation functions $F_2^{(1)}(\lambda)$ and $F_2^{(2)}(\lambda)$ in Eq.~(\ref{eq:jloop}) are
obtained as

\begin{eqnarray}
\label{eq:e56}
F_2^{(1)}(\lambda)=\frac{1}{\lambda}\,\frac{d}{d\lambda} F^{(0)}_2(\lambda)
 &=&\frac{g_A^4}{256\,\pi^3}\,\frac{\left(2\, m_\pi\right)^4}{f_{\pi}^4} \bigg\{
  {\rm e}^{-\lambda} \left(\frac{1}{\lambda^3}+\frac{1}{\lambda^2}\right) \left[1+\frac{1}{2}\int_0^1dz\, \frac{1-{\rm e}^{-\lambda(\alpha_z-1)}}{\alpha_z^2-1} \right] \nonumber\\
  &&\hspace{-0.25cm}- \left[ \left( \frac{1}{g_A^2}+2\right)
 \left(\frac{1}{\lambda^3}+\frac{1}{\lambda^2}\right) 
-\frac{1}{2\,\lambda}\right] \! E_1^{(-)}(\lambda)
- \left[ \left( \frac{1}{g_A^2}+2\right) 
\left(\frac{1}{\lambda^3}-\frac{1}{\lambda^2}\right)-\frac{1}{2\,\lambda}\right] 
\! E_1^{(+)}(\lambda)\nonumber\\
&&\hspace{-0.25cm}+ \left(\frac{1}{g_A^2} -2\right)\,\frac{1}{\lambda^3}\, E_1^{(0)}(\lambda)
-\int_0^1 dz \left(\frac{4}{\lambda^3}+\frac{1}{\alpha_z+1}\, \frac{1}{2\, \lambda^2} \right)
 {\rm e}^{-\lambda\alpha_z}\bigg\}\ , 
 \end{eqnarray}
\vspace{-0.5cm}
\begin{eqnarray}
\label{eq:e57}
F_2^{(2)}(\lambda)&=& \frac{d^2}{d\lambda^2} F^{(0)}_2(\lambda)
-\frac{1}{\lambda}\,\frac{d}{d\lambda} F^{(0)}_2(\lambda) 
= -\frac{g_A^4}{256\,\pi^3}\,\frac{\left(2\, m_\pi\right)^4}{f_{\pi}^4} \bigg\{
  {\rm e}^{-\lambda} \left(\frac{3}{\lambda^3}+\frac{3}{\lambda^2}+\frac{1}{\lambda}\right) \left[1+\frac{1}{2}\int_0^1dz\, \frac{1-{\rm e}^{-\lambda(\alpha_z-1)}}{\alpha_z^2-1} \right] \nonumber\\
&&\hspace{3cm}- \left[ \left( \frac{1}{g_A^2}+2\right)
 \left(\frac{3}{\lambda^3}+\frac{3}{\lambda^2}+\frac{1}{\lambda}\right) 
-\frac{1}{2}\left(1+ \frac{1}{\lambda}\right)\right] \! E_1^{(-)}(\lambda) \nonumber\\
&&\hspace{3cm}-\left[ \left( \frac{1}{g_A^2}+2\right)
 \left(\frac{3}{\lambda^3}-\frac{3}{\lambda^2}+\frac{1}{\lambda}\right) 
+\frac{1}{2}\left(1- \frac{1}{\lambda}\right)\right]   \! E_1^{(+)}(\lambda)
+ \left(\frac{1}{g_A^2} -2\right)\frac{3}{\lambda^3} \, E_1^{(0)}(\lambda) \nonumber\\
&&\hspace{3cm}-\int_0^1 dz \left[\frac{16}{\lambda^3}
+\left( 4\,\alpha_z+\frac{3}{2}\, \frac{1}{\alpha_z+1}\right) 
\frac{1}{\lambda^2}\right]
{\rm e}^{-\lambda\alpha_z} \bigg\}\ .
\end{eqnarray}
\end{document}